\documentclass[preprint2]{proto}
\usepackage{times,graphicx}
\newcommand{\refs}{\par\noindent\hangindent=1pc\hangafter=1}
\newcommand{\Msun} {M$_\odot$}
\newcommand{\Lsun} {L$_\odot$}

\newcommand{\um} {$\mu$m}

\newcommand{\simless}{\mathbin{\lower 3pt\hbox
      {$\rlap{\raise 5pt\hbox{$\char'074$}}\mathchar"7218$}}}
\newcommand{\simgreat}{\mathbin{\lower 3pt\hbox
     {$\rlap{\raise 5pt\hbox{$\char'076$}}\mathchar"7218$}}}

\begin{document}

\sloppy

\hyphenation {ISM}
\hyphenation {star}
\hyphenation {Sto-kes}
\hyphenation {spa-ti-al}
\hyphenation {spa-ti-al-ly}
\hyphenation {scat-te-ring}
\hyphenation {scat-te-red}
\hyphenation {nano-diamonds}

\title{\textbf{\LARGE Dust in Proto-Planetary Disks:\\ Properties and Evolution}}

\author {\textbf{\large Antonella Natta and Leonardo Testi}}
\affil{\small\em Osservatorio Astrofisico di Arcetri}

\author {\textbf{\large Nuria Calvet}}
\affil{\small\em University of Michigan}

\author {\textbf{\large Thomas Henning}}
\affil{\small\em Max Planck Institute for Astronomy, Heidelberg}

\author {\textbf{\large Rens Waters}}
\affil{\small\em University of Amsterdam and  Catholic University of Leuven}

\author {\textbf{\large David Wilner}}
\affil{\small\em Harvard--Smithsonian Center for Astrophysics}

\begin{abstract}
%\begin{list}{ } {\rightmargin 1in}
%%{\leftmargin 0in}
%\baselineskip = 11pt
%%rule{4.75in}{0.5pt}
%%\vskip 1pt
%\parindent=1pc
\baselineskip = 11pt
\leftskip = 0.65in 
\rightskip = 0.65in
\parindent=1pc

{\small 
We review the properties of dust in protoplanetary disks around
optically visible pre-main sequence stars obtained with a
variety of observational techniques, from measurements of scattered
light at visual and infrared wavelengths to mid-infrared spectroscopy
and millimeter interferometry. A general result is that grains in disks
are on average much larger than
in the diffuse interstellar medium (ISM). In many disks, there is
evidence that a large mass of dust is in grains with
millimeter and centimeter sizes, more similar to
``sand and pebbles" than to grains. Smaller grains (with micron-sizes)
exist closer to the disk surface, which also contains  much
smaller particles, e.g., polycyclic aromatic hydrocarbons.
There is some evidence of a vertical stratification, with smaller
grains closer to the surface. Another difference with ISM
is the higher fraction of crystalline relative to amorphous silicates
found in disk surfaces. There is a large scatter in dust
properties among different sources, but no evidence of  correlation 
with the stellar properties, for samples that include objects from
intermediate to  solar mass stars and brown dwarfs. There is also no 
apparent correlation
with the age of the central object, over a range  roughly between
1 and 10 Myr. This suggests a scenario where significant grain
processing may occur very early in the disk evolution, possibly when it is
accreting matter from the parental molecular core. Further evolution
may occur, but not necessarily rapidly, since 
we have  evidence  that large amounts of grains, from micron
to centimeter size, can survive  for periods as long as 
10 Myr.
 \\~\\~\\~}%leave this in to get the correct vertical space after the abstract

%\end{list}
\end{abstract}

\section{\textbf{INTRODUCTION}}

Young stars are surrounded by circumstellar disks, made of gas and dust.
Some of these disks will form planets, and one of the open questions
of the present time is to understand 
which disks will do it,  which  will not, and why.
In this chapter, we will refer to  all circumstellar disks
as {\em protoplanetary}, even if  some (many)
may only have the potential to evolve into a planetary system, 
and some (many) not even that.
 
The mass of protoplanetary disks  is dominated by gas. Still,
the solid component (dust grains) is of great importance. 
Observations of light
scattered on dust grains and their thermal emission remain key
diagnostic tools to detect disks and to characterize their
structure. 
Dust grains play an active role in determining the thermal and
geometrical structure of disks because their opacity dominates
over the gas opacity whenever  they are present. Furthermore,
grains shield the   disk midplane from energetic radiation,
thereby influencing the ionization structure,
possibly leading to a ``dead''  zone where the
magneto-rotational instability cannot operate. The formation of
dust grains is a phase transition which provides solid surfaces
important for chemical reactions and the freeze-out of molecular
components such as CO and H$_2$O in the colder parts of disks.
Finally, the solid mass of the disks is important because the dust
grains are the building blocks for the formation within the disk
of planetesimals and eventually planets.

Grains in protoplanetary disks are very different
from  grains in the diffuse interstellar medium;
their properties change from object to object and we believe
that in each object they evolve with time.
Without a comprehensive understanding of dust properties and
evolution, many disk properties cannot be understood.
An immediate example is 
the determination of disk solid masses from
millimeter continuum emission, which requires knowledge of the dust
absorption coefficient for a specific
grain composition and structure, and of the grain temperature.

Protoplanetary disks interest us also because they are the place
where planets form.
Grains are thought to be the
primary building blocks on the road to planet formation, and 
the continuing and growing  interest in this field is 
largely motivated by the need to understand
the diversity of
extrasolar planetary systems and reasons for this diversity.
Earlier
reviews by {\em Weidenschilling and Cuzzi} (1993) and {\em Beckwith et al.}
(2000) in {\em Protostars and Planets III and IV} summarize
what was known at the time.
Since {\em Protostars and Planets IV} we have seen a lot of progress in 
our  
knowledge of grain properties in disks, thanks to high-resolution data provided 
by infrared long-baseline interferometry, the high sensitivity of Spitzer and ground-based
10 m-class telescopes, millimetre interferometers such as PdB and OVRO,
and the VLA capabilities at 7 mm wavelength. 
Data exist now for a large number of disks, around
stars with very different mass and luminosity, from intermediate-mass
objects (Herbig AeBe stars, or HAeBe in the following)
to T Tauri stars (TTS) and brown dwarfs (BDs).
On the theoretical side,  there is
a revived effort  in 
modeling grain processing in disks (via coalescence, 
sedimentation, fragmentation, annealing etc.) and  its 
relation with the chemical and dynamical evolution
of the gas.
Although it is still very difficult
to  integrate observations and theory in a quantitative description
of disk evolution,  we can expect major advances in the near future.

In this review, we concentrate on observational evidence for
grain evolution with a special emphasis on grain growth and mineralogy.
As already mentioned,
grain properties and their distribution within the disk affect
many observable quantities.
These aspects will be covered in other
chapters of this book, and we will
concentrate on observations that measure directly the grain
physical and chemical structure.

Prior to that, we will review very briefly the most important
processes that control the grain properties (Section~2). We will then
outline the various observational techniques and their limitations
(Section~3), discuss evidence for grains of different size, from
Polycyclic Aromatic Hydrocarbons (PAHs; Section~4) to micron-size
(Section~5) and centimeter-size (Section~6). We will discuss grain
mineralogy in Section~7.  Section~8 will summarize the main conclusions one
can derive from the observations and outline some of the open
questions we need to address in the future.

%In general, information on grain sizes comes from an analysis of
%extinction curves, scattering and polarization data, the
%shape/presence of dust features, and thermal continuum emission,
%especially at millimeter wavelengths where the disks become
%optically thin. The detailed investigation of infrared bands,
%mainly coming from silicates and other oxides, provide insight in
%the chemical and structural evolution of the solid component in
%protoplanetary disks. 

\section{\textbf{GRAIN GROWTH: WHY AND HOW}}

Grains in the diffuse interstellar medium (ISM) are very likely
a mixture of silicates and carbons, with  a size distribution from
$\sim 100$ \AA\ to maximum radii of $\sim 0.2-0.3$ \um. Smaller
particles, most likely PAHs and very
small carbonaceous grains, are also present (e.g., {\em Draine}, 2003).
The composition of  dust in molecular clouds is similar. Models of
dust evolution in  collapsing cores  ({\em Kr\"ugel and Siebenmorgen}, 1994;
{\em Ossenkopf and Henning}, 1994;
{\em Miyake and Nakagawa}, 1995) 
predict only minor changes, as confirmed 
by the observations of Class 0 and Class I objects (e.g., 
{\em Beckwith and Sargent}, 1991;
{\em Bianchi et al.}, 2003; {\em Kessler-Silacci et al.}, 2005).

%Before we discuss the observational data, we  want to
%summarize  why we expect grains to grow in disks, and
%how we think this may happen. A deeper discussion can be found
%introduce shortly the main physical processes expected to operate
%in dust grain evolution. For the discussion of the various theoretical scenarios
%we refer to the contribution by
%in {\em Dominik et al.} in this volume.
Major changes occur once dust is collected in a circumstellar disk,
where the  pristine interstellar grains may grow  from sub-micron size or even
smaller particles, to  kilometer-sized bodies (planetesimals), and eventually planets. 
This process is driven by coagulation of smaller particles into larger
and larger ones, and has a long history of theoretical and laboratory studies,
which it is  not the scope of this paper to analyze. We refer for a critical
discussion to the most recent reviews of {\em Henning et al.} (2005)
and the chapter by {\em Dominik et al.} 
However, the relevance of
the observations of grain properties we will discuss in the following is
better understood
in the context of grain growth and planetesimal formation
models, which we will therefore briefly summarize here.

%The dense and cold environment found
%in protoplanetary disks is favorable to grain growth by collisional
%coagulation. This process is strongly enhanced by the sedimentation
%of grains toward the disk midplane, under the effect of stellar gravity.
%The efficiency of this process depends on a number
%of poorly known conditions. 
%for example the velocity at which grains
%collide, or the probability that collisions of different kind
%lead to fragmentation rather than coalescence (see, e.g., {\em
%Chokshi et al.}, 1993). 
%In addition, mixing processes, both in the vertical and in the
%radial direction, may affect grain growth, in ways that
%are not yet understood. The most recent reviews of these processes
%are {\em Henning et al.} (2005) and {\em Dominik et al., this volume}.
Grain growth to centimeter- and even meter-sizes
is mainly driven by collisional aggregation, although
gravitational instabilities may  play a role in
over-dense dust regions. The relative velocities leading to such
collisions are caused by the differential coupling to the gas
motion (see, e.g., {\em Weidenschilling and Cuzzi}, 1993; {\em
Beckwith et al.},  2000). This immediately demonstrates that grain
growth cannot be understood without a better characterization of
the gas velocity field in disks, especially the degree of turbulence.
For sub-micron-sized grains this is not a real issue and their
growth by Brownian motion is reasonably well understood thanks to
numerical simulations and extensive laboratory experiments (see,
e.g., {\em Blum},  2004 and the chapter by {\em Dominik et
al.}). The relative velocities of grains in the
1-100 $\mu$m size range are of the order of 10$^{-3}$ m/s to
10$^{-4}$ m/s, low enough for sticking. The outcome of this early
growth process is a  relatively narrow  mass
distribution function with fractal aggregates having open and
chain-like structures. 
%The mass increases in time as t$^2$ (see, {\em Blum}, 2004). 

While growing in mass, particles also start to sediment
(due to the vertical component of the stellar gravitational field,
{\em Schr\"apler and Henning}, 2004, {\em Dullemond and Dominik}, 2005 and references therein)
 and the relative velocities may eventually
become higher leading to the compaction of aggregates. Above the
compaction limit, a runaway growth can be expected, where a few
large aggregates grow by collisions with smaller particles. In
this regime an exponential increase of mass with time can be
expected. The coupling between sedimentation, grain growth and
structure of particles remains the next challenge for
protoplanetary dust models which have to deal with the transition
from open to compact aggregates. For grains of boulder size
(larger than 1 m), the
relative velocities remain more or less constant at about 50 m/s,
challenging growth models because this velocity is above the
destruction threshold (but see {\em Wurm et al.}, 2005). Furthermore, such boulders are rapidly
transported to the central star within a few hundred years. There
are quite a number of proposals to solve the problem of the growth
barrier, none with complete success yet.

In spite of the many uncertainties, the models converge on predicting
that grains can grow to very large sizes and sediment under the
effect of the stellar gravity, leaving
a population of smaller grains closer to the disk surface and
of increasingly larger bodies toward the midplane. Grain growth
should be much slower in regions further away from the
star  than in regions closer to it,
which should be completely depleted of small grains.
When only coagulation mechanisms are considered, 
small grains are removed very quickly,
leading to the complete disappearance of the dusty disks on timescales
much shorter than the age of observed disks. Mechanisms that
replenish  disks of small grains are clearly required, for
example aggregate fragmentation. Radial and meridian
circulation can also play an important role in grain evolution. 

The  process  of growth of grains from  submicron to km-size we have described
assumes that the  original solid mass  never  returns to the gas phase. This may not be the case, as
the temperatures in the inner disk are high enough to evaporate 
grains. If, over time,
radial circulations  carry close to the star
most of the disk material,   all
the original solid mass is destroyed and condensed again in the disk.
If so, the newly formed grains may be very different from those in the
ISM, even if actual growth does not occur, or is important only above
a certain (large) size.
Gassification and recondensation of grains in the inner disk,
coupled with strong radial drifts, has been suggested ({\em Gail}, 2004)
as an explanation for
the large fraction of crystalline grains seen in several disks (see Section~7),
and may play an important role also in determining the grain size distribution.
%However, in the following we
%will concentrate on
%``growth"  when  discussing   grains larger than the ISM ones.

For a long time, the only test to planet formation theories was 
our own Solar system. This has changed completely over the last decade,
first with the discovery of the variety of planetary
systems that can actually form, but also with a much better knowledge
of the grain properties and distribution in disks which are
the progenitors of planetary systems. Although our knowledge
and understanding of this second aspect is still limited,
and the answers to some key questions (e.g., which planetary
systems will form and from which disks) elusive,
we will show in the following that the
observations  may already provide important constraints to theory.

\section{\textbf{OBSERVATIONAL TECHNIQUES}}
\label{sobs}

Grain properties in disks can be explored with a variety of techniques
and over a large range of wavelengths, from visual to centimeter.
Before reviewing the results and their implication, it is useful to
summarize briefly the capabilities and limitations of the
different techniques.

The most severe limits to the characterization of grains in disks do not
come 
from observational limits, such as spatial resolution or sensitivity
(although these too can be a problem), but from the  physical
structure of disks. Their extremely high
optical depth allows us to  get information only on grains 
located in a $\tau=1$ layer, which at optical
and IR wavelengths contains a tiny fraction of
the dust mass. Moreover, if one  measures dust emission, the
observed flux is strongly biased by the temperature dependence of the
Planck function.

The only way to measure the properties of the bulk of the
dust mass is to go to  longer wavelengths, where an increasing fraction of
the disk becomes optically thin. Interferometric observations at millimeter and
centimeter wavelengths have provided  the strongest evidence so far that
most of the original solid mass in  protoplanetary
disks  has grown to cm-size by the time the central star becomes
optically visible. However, one should remember that this technique is
limited today by the sensitivity and resolution of  existing
millimeter interferometers. For example, studies of the
radial dependence of the dust properties are still very difficult.
Furthermore, even at these very long wavelengths, 
the regions of the disk closer to the star
(typically, up to few AUs) are optically thick and
thus this very interesting inner portion of the midplane remains inaccessible
at all wavelengths.
%Finally, there are no dust features at very long
%wavelenghts, so that the dust characterization is necessarily limited
%to deconvolving the dust opacity.

The mid-infrared spectral region, roughly between 3 and 100 $\mu$m, is
rich in vibrational resonances of abundant dust species, silicates in
particular. 
The spectral region between 3 and 20 $\mu$m
contains also prominent PAH C-C and C-H resonances. 
The wavelength, spectral shape and strength of the resonances are sensitive
probes of the chemical composition, lattice structure, size and shape
of the particles. In disks around optically visible
stars these dust features are generally seen in emission, having origin in the
optically thin surface layers, heated by the stellar radiation to
temperatures higher than the disk midplane (e.g., {\em Calvet et al.}, 1991;
{\em Chiang and Goldreich}, 1997; {\em Menshchikov and Henning}, 1997).
%Some dust species show several resonances covering a
%wide wavelength range, making them also probes of the temperature
%(distribution) of specific grain types in the circumstellar
%environment. 
Infrared spectroscopy roughly probes dust particles with
sizes of up to a few microns (depending on material and
wavelength), in the temperature range between 1500 and about 50~K. In
protoplanetary disks, this implies that such spectra are sensitive to
dust in the innermost regions, and in the surface layers of the disk only.
No information is obtained about chemical and allotropic
properties of large and/or  cold dust grains.
%Exceptions are the spectra of edge-on disks where volatiles (e.g.
%CO$_2$, H$_2$O, or CO ice) are seen in absorption ({\bf ref van Dishoeck}). 
%These ice mantles play a very important role in the
%efficiency of grain growth ({\bf ref??}).
%In very few cases dust features are seen in absorption. This can
%happen in actively accreting disks, where the temperature is higher
%in the midplane than in the surface (e.g., ??). 

The 10 $\mu$m spectral region deserves special attention because
it is accessible from the ground, is available for high spatial
resolution interferometric studies, and a fairly large body of data has
been collected, covering a wide range of stellar mass and age. It
contains the strongest resonances of both amorphous and crystalline
silicates. The
silicate emission bands have important but somewhat limited diagnostic
value. First, the 10 $\mu$m spectral region probes grains of a certain
temperature range, mostly between 200 and 600~K 
%(although contributions from a wider temperature range are also possible). 
Second, both the
amorphous and crystalline silicate resonances near 10 $\mu$m only probe
presence  of grains smaller than a few microns in size, the exact value
depending on the dust species. This limits the diagnostic value of the
10 $\mu$m region to warm, small silicates mostly located on the disk
surface in 
regions within  $\sim$ 10~AU for  protoplanetary disks
around HAe stars, $\sim 0.5-1$ AU for TTS, and
$\sim 0.1$ AU for BDs.

At shorter wavelengths (visual and near-IR), scattering becomes
important and one can observe scattered
light extending to very large distances from the star (e.g., {\em McCaughrean et al.}, 2000).  
Multi-wavelength images in scattered light  can be used to derive grain
properties over a large range of disk radii, but limited to a very narrow
layer on the disk surface.

Interesting information on grain properties deeper in the disk can be
obtained from edge-on disks (silhouette),
seen as a dark lane against a luminous background.
Information on dust properties can be
obtained in favorable cases by studying the wavelength dependence of the
dark lane absorption against the background of the scattered light
at visual and near-infrared wavelengths.
Silhouette disks should have
deep silicate features at 10 \um\ in absorption, which,
contrary to the features in emission, sample
the colder grains in the outer disk. 
Silicate absorption is observed in few cases, but it is often
difficult to  exclude contamination from grains in the surrounding
material. 
%{\bf Orion Proplyds?}. 

Multiwavelength studies are obviously necessary to completely characterize 
the grain population, in size, composition and physical structure.
So far, this cannot be done in disks, where observations at different wavelengths
sample different regions of the disk.
As we will describe in the following,
there is clear evidence that grains of all sizes (from PAHs to
cm-size bodies at least) are 
present in disks.
However, the information is always partial, and it is very difficult to get a global picture, for example, to measure the total mass fraction of grains of different size.

An additional caveat is that none of the techniques we have mentioned is sensitive
to bodies larger than few centimeters. Kilometer-size planetesimals
can be detected through the dynamical perturbations they create, but
the detection of meter-size bodies is practically impossible.

\section{\textbf{
THE SMALLEST PARTICLES: PAHS
}}

Emission from transiently heated very small particles
(at 3.3, 6.2, 7.7-7.9, 8.6, 11.3 and 12.7 $\mu$m etc.) 
has been detected in many
disk systems, mostly HAeBe stars (e.g., {\em Acke and van den Ancker},
2004) and generally attributed to PAHs.
%but also
%a few objects of  spectral type F (e.g., {\em Sloan et al.}, 2005). 
For a long time,
it has not been clear if the observed emission is from PAHs in the disks
or from the reflection nebulosities associated with many
HAeBe. There is now convincing evidence in favour of the
disk origin. 

{\em Peeters et al.} (2002),  {\em van Diedenhoven et al.} (2004)
and {\em Sloan et al.} (2005)
have shown that in HAeBe stars 
the PAH features differ in shape and wavelength when compared to the
ISM.
{\em Meeus et al.} (2001) and {\em Acke and van den Ancker}
(2004) found that the strength of the PAH bands
correlates with the shape of the spectrum in the 10-60 $\mu$m region.
Flared disks (i.e., with flux $F_\nu$ increasing with wavelength in the
10-60 $\mu$m range; group I in the {\em Meeus et al.} (2001) 
classification) have strong PAHs features,
while flat disks (i.e., with $F_\nu$ decreasing with
wavelength in the 10-60 $\mu$m range;
group II) have no or very weak emission.
This trend has been analyzed by {\em Habart et al.} (2004a), who have
computed disk models which include PAHs and 
shown that the observed
correlation between the strength of the PAH
bands and the shape of the SED is well explained in terms of the
solid angle the disk subtends as seen from the star.
%, and the number of
%PAH molecules in the disk surface that can be excited by stellar
%photons.
In a few stars, other features, at 
3.43, 3.53 $\mu$m, from transiently heated
carbonaceous materials, possibly identified as nanodiamonds
({\em Guillois et al.}, 1999), have also been  shown to have a disk origin 
({\em Van Kerckhoven et al.}, 2002).

\begin{figure}[ht]
 \epsscale{1.0}
%\plotone {ps.naco}
\plotone {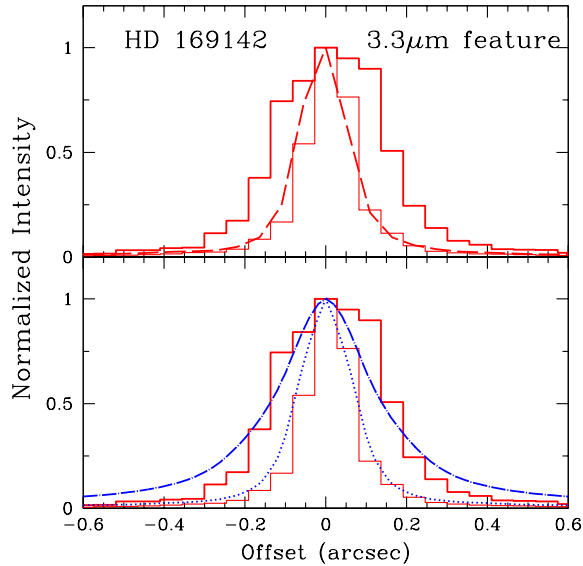}
\caption{Intensity profile of the 3.3 \um\ PAH feature in the HAe star
HD~169142, observed with NAOS/CONICA on the VLT ({\em Habart et al.}, 2006). 
The top panel shows the intensity in the feature after
continuum subtraction (thick solid line) and
in the adjacent continuum (thin solid line); the dashed line plots
the point-spread function as measured on a nearby unresolved star.
The PAH feature is clearly spatially resolved,
with FWHM size of 0.3 arcsec (roughly 40 AU at the distance of the
star). The continuum is unresolved.
The bottom panel compares the observations
with disk model predictions for the feature (dot-dashed line) and
the continuum (dotted line). Details of 
observations and calculations, including model parameters,
are in {\em Habart et al.} (2004a, 2006).
}
\label{naco}
\end{figure}

Recently, PAH and nanodiamond emission  has been spatially
resolved in several objects 
({\em van Boekel et al.}, 2004b; {\em Ressler and Barsony}, 2003;
{\em Habart et al.}, 2004b, 2006);
the emitting region has a size  consistent with disk
model predictions ({\em Habart et al.}, 2004b, 2006; see Fig.~\ref{naco}).
%Because the emission in the features is more extended
%than the adjacent continuum (due to larger grains in thermal
%equilibrium with the stellar radiation field) 
The emission of such transiently heated species (which are
``hot" whenever excited) is much more
extended than that of the adjacent continuum, emitted by
larger grains in thermal equilibrium with the radiation field
which become rapidly cool as the distance from the star increases.
Using images in the  features of PAHs and nanodiamonds
it has been possible 
to measure disk inclination and position angle on the sky 
({\em Habart et al.}, 2004b)
at spatial scales of few tens AU.

The presence of very small particles on disk surfaces has a strong
impact on the gas physical properties, since they contribute a
large fraction of the gas heating via the photo-electric effect
and may dominate the H$_2$ formation on grain  surfaces.
PAHs are thus an essential ingredient of
disk gas models (e.g. chapter by {\em Dullemond et al.}).
However, the observations have severe limitations, as
they can only probe matter at large altitude above the
disk midplane, so that we do not know if PAHs 
survive deeper into the disk; moreover, the intensity in the
features decreases with radius , and
is well below current observational capabilities for
distances $>$40 AU (at least for the 3.3 $\mu$m feature).

There are also limitations in the current models, which need to explore a 
much wider range of parameters, both in  disk and PAH
properties. 
%For example,
%the Habart et al. (2004) models  can roughly reproduce
%the observations assuming that on the disk surface
%PAHs are present in a proportion (relative to larger grains) similar
%to what is measured in the ISM.

% have classified the ISO spectra of Herbig Ae/Be
%stars in two groups, group I containing red, rising spectra in the
%10-60 $\mu$m range, while group II has bluer, declining spectral
%shapes. This classification is consistent with flaring (group~I) and
%flat (group~II) disks. Theoretical models for passively heated,
%\emph{self-shadowed} disks have given this phenomenological
%classification a firm basis (Dullemond 2003; Dullemond \& Dominik
%2004). Group~I sources are disks in which the outer regions emerge from
%the shadow of a puffed-up inner rim (Dullemond et al. 2001), while
%in group~II sources the outer disk never emerges from
%the inner disk shadow. 

\section{\textbf{
FROM SUB-MICRON TO MICRON GRAINS 
}}

\bigskip
\noindent
\textbf{5.1
Scattering and Polarization}

%Astronomers are used to the fact that the wavelength dependence of
%interstellar extinction can be used to constrain the size of dust
%grains. Absorption dominates over scattering for particles
%sufficiently small compared with the wavelength. If extinction is
%dominated by absorption then the extinction efficiency will vary
%as 1/$\lambda$, otherwise we expect a 1/$\lambda^4$ behaviour.
%The almost linear decrease of interstellar extinction with
%wavelength immediately implies that most of the particles should
%have sizes smaller than the wavelength of visible light.

Observations at visible and near-infrared
wavelengths of the dark disk silhouette seen
against the background light scattered by the disk surface 
provide direct view of disks around
young stars. 
The properties of
the scattered radiation and  the
fraction of polarization can be used in principle  to measure the size of the
scattering grains.
%Observations of protoplanetary disk at visible and near-infrared
%are often only possible because we see light scattered off the
%disk surface or the disk environment. 
%In fact, in the laboratory
%angular scattering is more frequently used to determine particle
%sizes than just extinction. 
However, we know from laboratory work that particle sizing
based on the angular distribution of scattered light or the
measurement of polarization is not without problems and usually
only works well for monodisperse distributions of spheres. In 
astronomical sources,
the analysis of scattered radiation and polarization is
even more complicated because it is not possible to obtain measurements
for  all scattering angles. 
In addition, polarization
measurements are often technically challenging which may explain
why we only have limited data using this technique and grain size
information from such observations is still scarce. 
%For comprehensive
%studies of objects in the Taurus-Auriga region, using scattering 
%(and polarization), we refer to {\em Whitney et al.}, 1997, {\em Lucas
%et al.},  2004, {\em Duch\^{e}ne et al.}, 2004). 

Scattering is well described by the 4x4 scattering matrix (Mueller
matrix) transforming the original set of Stokes parameters into a
new set after the scattering event. The matrix elements are
angular-dependent functions of wavelength, particle size, shape,
and material. For particles that are small compared with the wavelengths,
the scattered light is partially polarized 
%if the incident light is
%unpolarized 
with a typical bell-shaped angular dependence with
100\% polarization at a scattering angle of 90 degrees. The
polarization degree depends only on the scattering angle and not
on particle size. This practically means that the radius of small
spherical particles cannot be determined from scattering
measurements.

For very small particles scattering does not vary a lot with
direction. This changes drastically for larger spheres with
more ripples and peaks in the scattering pattern, for which forward
scattering becomes very important. In addition, the peak
polarization decreases and moves to larger scattering angles. The
very structured scattering pattern for large spheres is used for
experimental particle sizing, but the characteristic ripples
disappear if a grain size distribution is considered instead of a single 
size population. For very large spheres we
again find smooth curves. Furthermore, non-spherical particles may
behave very differently. We should also note that an ensemble of
non-identical scattering particles leads to depolarization of
scattered light, adding to the complexity of our understanding of
the polarization degree (in terms of grains sizes) of radiation
coming from a disk. We refer to {\em Voshchinnikov and Kr\"ugel} (1999)
for Mie calculations, demonstrating the effect of different grain
sizes. 
%\textbf{Should we add there Fig.3 showing the effects nicely?}

One may conclude that solving the inverse problem of
determining grain sizes from scattering and polarization
observations is hopeless. This is certainly not true.
%especially because the degree of polarization, the different
%angular scattering patterns and the wavelength dependence of
%polarization can be used to trace grain size evolution. 
A growing number of objects are studied using multi-wavelength
imaging and polarization techniques.
For our purposes, it is sufficient to note that, 
in all cases,
the evidence points toward
grain growth on the disk surface  to 
sizes  of up to a few microns; often, there is also evidence of sedimentation, i.e.,
that larger grains are closer to the disk midplane. Among
the more recent studies, we note
%Comparing simulated near-infrared aperture polarimetric data 
%of disk+envelope systems with
%observations, Fischer et al. (1996) concluded that grains sizes
%have to be larger than in the ISM, at least in the disk,
%otherwise the polarization degree would be too high. 
%%However, one
%should note that the polarization degree observed in the outer
%nebulosities around YSOs in Taurus is indeed often very high (e.g.
%Whitney et al. 1997), indicating that small grains dominate the
%scattering cross sections.  
%Voshchinnikov \& Kr\"ugel (1999)
%analysed the available polarization data of the edge-on disc
%around $\beta$ Pic in the BVRI-bands with spherical particels and
%a power-law size distribution. They concluded that the minimum
%size of the dust particles is larger than in the general
%interstellar medium with a$_{min}$=0.15 $\mu$m and with an
%ill-defined maximum size. Krivova et al. (2000) improved the size
%distribution and investigated the asymmetry in the polarization of
%the two wings separately. 
the work of {\em Lucas et al.} (2004), who nicely demonstrated
the power of a detailed analysis of high-resolution imaging
polarization data for the case of HL Tau. They found that silicate
core-ice mantle grains with the largest particles having radii
slightly in excess of 1 $\mu$m best fit the data. A similar study
has been performed for the ring around GG Tau by {\em Duch\^{e}ne et al.} 
(2004).
They found again a slight increase in the grain size towards micron-sized
grains, and evidence  for a vertical stratification of dust.
 The surface layers, located ~50 AU above the ring midplane, contain dust 
grains that are consistent with being as small as in the ISM, 
while the region of the ring located ~25 AU from the midplane contains 
significantly larger grains ($>$ 1 $\mu$m). 
This stratified structure is likely the 
result of vertical dust settling and/or preferential grain growth in the 
densest parts of the ring.
The first scattering measurements at mid-infrared wavelengths obtained for
the T Tauri star HK Tau B ({\em McCabe et al.}, 2003) showed similar results.

Evidence of growth of grains to sizes
of up to a few microns and of a vertical stratification
of grain sizes within the disk surface has
also been inferred in silhouette disks 
in Taurus 
({\em D'Alessio et al.}, 2001) and Orion ({\em Throop et al.}, 2001;
{\em Shuping et al.}, 2003; see also {\em McCaughrean et al.}, 2000 and
references therein),
by studying the behaviour with wavelength of
the translucent edges of the dark lane. 
%The width of the dark lane is a sensitive function of the 
%grain size and fits better the observations with larger grains.
%The decreasing width of the dark
%lane with larger grains is practically the result of a decreased
%opacity. 
%In their calculations  a change
%of the scattering properties due to the presence of larger grains
%has not been considered.

\bigskip
\noindent
\textbf{5.2
Mid-Infrared Spectroscopy}

High quality mid-IR spectra are available for an increasing number of
objects.
% especially in the 10 $\mu$ region, but also at longer wavelengths. 
First ISO and now Spitzer are providing us
with data of a quality that could not be achieved from the ground, even
if the spatial resolution  is in both cases too low to resolve
the emission. 
%Among the  most recent results, there are
%the ground-based spectra of HAe stars by
%{\em van Boekel et al.} (2004);
%% and {\em Acke and van den Ancker} (2004);
%the
%TTS  ground-based spectra
%of {\em Meeus et al.} (2001, 2003), {\em Hoda et al.} (2003),
%{\em Przygodda et al.} (2003),
%{\em Kessler-Silacci et al.} (2005b), and  the 
%Spitzer data from the legacy programs c2d  ({\em Kessler-Silacci et al.}, 2005a)
%feps ({\em Bowman et al.}, 2005), and the GTO program ??.
%({\em Forrest et al.}, 2004, {\em Sargent et al.}, 2005).
%Spitzer spectra of 6 brown dwarfs in Chamaleon have been obtained
%by {\em Apai et al., (2005)}. 
Spitzer spectra are becoming available
in increasing number at the time we write, and  we expect further
improvement in the statistics and quality of the data, for low
mass objects in particular.
In the following, we provide a summary of the silicate
properties, as derived from the available observations.

\begin{figure*}[t]
 \epsscale{1.0}
%\plotone {ps.sil_profiles}
\plotone {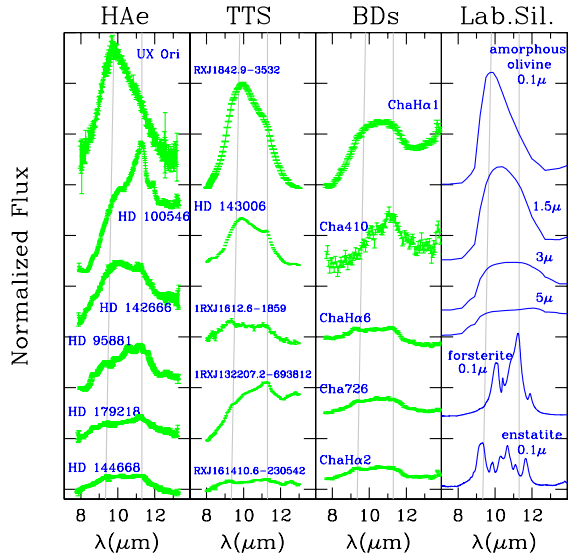}
%\includegraphics[width=15cm]{natta_fig2.ps}
%  \begin{center}
%    \leavevmode
%    \centerline{ \psfig{file=ps.sil_profiles,width=10cm,angle=0} }
%  \end{center}
\caption{Observed profiles of the 10 \um\ silicate feature.
The first three Panel (from left
to right)   show a selection of profiles for HAe stars
({\em van Boekel et al.}, 2005), 
 TTS ({\em Bouwman et al.}, 2006), 
and  BDs in Chamaleon ({\em Apai et al.}, 2005). 
All the profiles have been normalized to the 8 \um\ flux and shifted
for an easier display. In each Panel, they have been ordered according
to the $F_{peak}/F_{cont}$ value. The dashed vertical lines show the
location (at 9.8 \um) of the peak of small amorphous  olivine 
(Mg$_2$SiO$_4$)
and of the strongest crystalline feature of forsterite at 11.3 \um .
Amorphous silicate grains of composition
Mg$_{2x}$Fe$_{2-2x}$SiO$_4$, generally with $x\ll 1$ have the same composition
of olivines, and are are often referred to as
``amorphous olivines". Although this is not correct (olivine is crystalline), 
we will sometime use this definition. 
The Panel to the extreme right shows for comparison a selection of
profiles of laboratory silicates. Starting from the top,
amorphous olivine with radius 0.1, 1.5, 3 and 5 \um, as labelled. In all cases,
we have added the same continuum (30\% of the peak
of the 0.1 \um\ amorphous olivine), and normalized as the observed profiles.
The two bottom curves show the profile of small crystalline
grains, forsterite (Mg$_2$SiO$_4$) and enstatite (MgSiO$_3$), respectively; 
larger crystalline grains
have broader and weaker features (e.g., {\em van Boekel et al.}, 2005).
Several of the features displayed by crystalline silicates can be identified
in some of the observed spectra.
}
\label {sil_prof}
\end{figure*}

Most of the  information is obtained from observations
of the 10 \um\ emission feature.
The  spectra show 
a large variation in the strength and shape of this
feature, for objects of all mass,
from HAe stars to TTS and BDs.
Fig.~\ref{sil_prof} plots a representative selection of  
HAe stars, TTS, and BDs, which shows the variety of observed profiles.
In some cases, the shape of the feature is strongly peaked at about
9.8 \um, as for small ($\ll 1$ \um) amorphous grains in the ISM.
In other objects, the feature is very weak with respect to the continuum,
and much less peaked.
In other cases, many narrow features, typical of
crystalline silicates, are clearly visible,  overimposed on the 
smoother and broader amorphous silicate emission.
These features will be discussed in detail in Section~7.
In spite of these large variations,
there is a general trend, valid for objects of all mass,
first
identified by {\em van Boekel et al.} (2005),
who noted that the shape and
strength of the silicate band are correlated, so that weaker features
are also flatter.  
%(see {\em van Boekel et al.
%2003; Meeus et al. 2003; Przygodda et al. 2003), 
This is consistent
with the growth of  silicate grains
from sub-micron   to  micron sizes.
If grains grow  further, the silicate emission disappears;
there are in fact a few  objects which show no  silicate emission, most
likely because \emph{all} small silicates have been removed (e.g. 
{\em Meeus et al.}, 2003).  

\begin{figure}[ht]
 \epsscale{1.0}
%\plotone{ps.sil_ratios}
\plotone {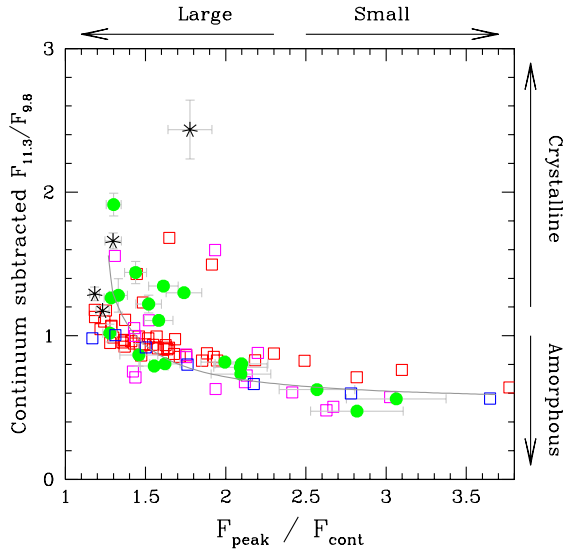}
\caption{ The ratio of the 
continuum-subtracted flux at 11.3 \um\ over that at 9.8 \um\
($F_{11.3}/F_{9.8}$) is plotted as function of
flux at the peak of the 10 \um\ feature over
the continuum, $F_{peak}/F_{cont}$. 
Open and filled symbols show the observations for
HAe stars ({\em van Boekel et al.}, 2005; filled dots),
TTS ({\em Kessler-Silacci et al.}, 2006, {\em Bouwman et al.}, 2006,
{\em Przygodda et al.}, 2003;
squares), BDs ({\em Apai et al.}, 2005; stars).
All the data but those of {\em Kessler-Silacci et al.} (2006)
and {\em Bouwman et al.} (2006) have been re-analyzed
by {\em Apai et al.} (2005) in a homogeneous way.
Error bars are shown when available in the literature.
The solid curve shows the result for a dust model which includes
amorphous olivine of two sizes  (1 and 10 \um)
in proportion varying from all 1 \um\ (extreme right)
to all 10 \um\ (extreme left), and an additional contribution
of a 2\% mass fraction of 0.1 \um\ forsterite grains
(from J. Bouwman, private communication); the same  continuum 
(25\% of the peak flux of the smallest grains) is added to all
model spectra.
}
\label {sil_ratios}
\end{figure}

A useful way to summarize the properties of the 10 \um\ feature of
a sample of objects is shown in
Fig.~\ref{sil_ratios}.
For each object,
the shape of the 10 \um\ silicate emission
feature is characterized by two parameters,
the ratio of the flux at the peak of the feature over the
continuum ($F_{peak}/F_{cont}$) and the ratio of the continuum
subtracted 11.3 \um\ over the 9.8 \um\ flux.
Amorphous silicates 
of increasing size have weaker and flatter
features (cf. Fig.~\ref{sil_prof}), with
values of $F_{peak}/F_{cont}$ decreasing from $3-4$ (for
sizes $\ll$1 \um) to $\sim 1$, for grains larger than few microns.
At the same time, the ratio
$F_{11.3}/F_{9.8}$ increases  from $\sim 0.5-0.6$ to $\sim 1$.
Larger values of the ratio $F_{11.3}/F_{9.8}$ are due to
a significant contribution from forsterite (i.e.,  crystalline 
Mg$_2$SiO$_4$).
%However, this does not seem to be very common in
%objects with mid-IR excess emission (??).

Note that the ratio $F_{peak}/F_{cont}$
depends not only on grain cross section but also on
other quantities (e.g., the disk inclination and geometry)
which affect differently feature and  continuum emission
(e.g., {\em Chiang and Goldreich}, 1999).
Also,  changes in   grain properties other than size (e.g., porosity; see
{\em Voshchinnikov et al.}, 2006, {\em Min et al.}, 2006)
can reproduce part of the  observed trend, and make the simple
analysis in terms of grain size more uncertain.
However, the interpretation of the 10 \um\ feature profiles in terms
of growth of the grains in the disk surface is convincing,
and in agreement with the results from other kinds of observations
discussed in Section~5.1.

%Fig.?? shows data for HAe, TTS and BDs. It is easy to see that the
%observed correlation is similar for objects of very
%different mass.

Additional information will be  derived in the future
from the properties of the
weaker 20 $\mu$m silicate band, which 
samples slightly larger grains, and regions
on the disk surface further away from the star.  
Spitzer spectra are becoming available for
 a rapidly growing number of objects, and the
first results seem to confirm the results obtained from the
10 $\mu$m feature  analysis ({\em Kessler-Silacci et al.}, 2006;
{\em Bouwman et al.}, 2006).

\section{\textbf{
GROWTH TO CM-SIZE GRAINS
}}

As discussed in Section~\ref{sobs}, at (sub-)millimeter wavelengths and
beyond the dust emission in disks begins to be moderately optically
thin. At these wavelengths it is thus possible to probe the bulk
of the dust particles, which are  concentrated in the disk midplane.
The spectral energy distribution of the continuum emission can be directly
related to the dust emissivity, which in turn is related to the grain properties.
The variation of the dust opacity coefficient per unit mass
with frequency, in the millimeter wave range, can be approximated to
relatively good accuracy by a power law: $k\sim\lambda^{-\beta}$.
The value of the $\beta$ exponent is directly related to the dust properties,
in particular to the grain size distribution and upper size cutoff;
% ({\em Miyake and Nakagawa}, 1993;{\em Beckwith et al.}, 2000; {\em Natta et al.}, 2004). 
in the limit
of dust composed only of grains much larger than the observing wavelength,
the opacity becomes grey and $\beta=0$.

It has been known for many years that pre-main sequence disks have (sub-)millimeter
 spectral indices
$\alpha$ ($F_\nu \propto \lambda^{-\alpha}$) shallower than 
prestellar cores and young protostars ({\em Beckwith and Sargent}, 1991). 
If the disk emission is optically thin and
$h\nu/{\rm k}T_{d}\ll 1$,
the observed values of $\alpha$ would translate very simply
in opacity power law indices $\beta=\alpha-2$; with observed typical values
of $\alpha\simless 3$, this would imply
$\beta \sim 1$ or lower. This immediately suggests 
that disk grains  are very different from ISM grains, which
have $\beta \sim 1.7$ ({\em Weingartner and Draine}, 2001).

This result, however, has been viewed with great caution,
since the effect of  large optical depth at the
frequency of the observations cannot be ruled out for
spatially unresolved observations. The extreme case of an optically thick
disk made of ISM grains has $\alpha=2$, the same value as an optically thin disk with
very large grains, for which $\beta=0$ (see, e.g.,
{\em Beckwith and Sargent}, 1991; {\em Beckwith et al.}, 2000). 
The combination of optical depth and grain properties can be
disentangled if
the emission is resolved spatially and the disk size at the observed
wavelength  is measured
(see, e.g., {\em Testi et al.}, 2001, 2003).

\begin{figure}[ht]
 \epsscale{1.0}
%\plotone{ps.beta_obs}
\plotone {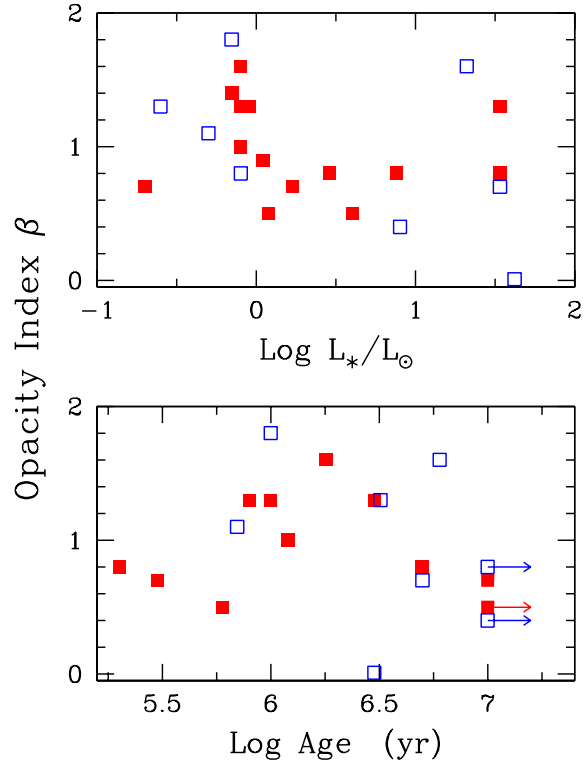}
%  \begin{center}
%    \leavevmode
%    \centerline{ \psfig{file=ps.beta_obs,width=7cm,angle=0} }
%  \end{center}
\caption{ The figure plots the values of the millimeter opacity
index $\beta$ ($\kappa \propto \lambda^{-\beta}$)
as function of the stellar luminosity (top
panel) and of the stellar age (bottom panel);
data
from {\em Natta et al.} (2004) and {\em Rodmann et al.} (2006). Filled squares are
objects where the disk millimeter emission has been spatially resolved,
open squares objects  which  have not been observed with high
spatial resolution, or that were found to be not resolved with the
VLA (resolution $\sim 0.5$ arcsec).
}
\label{beta_obs}
\end{figure}

There are now several disks whose millimeter emission has been
imaged with interferometers.
A major step forward has come from the use of 7 mm VLA data, which, in combination with results at shorter wavelengths (1.3 and 2.6 mm) with PdB and OVRO,
provides not only
 high spatial resolution and  larger
wavelength range (minimizing the uncertainties on $\alpha$), but
also the possibility of probing population of grains as large as a few centimeters.
Additionally,  the VLA at centimeter wavelengths has  been
useful to check whether the dust emission at shorter wavelength is 
contaminated by free-free emission  (e.g. {\em Testi et al.}, 2001; Wilner et al. 2000; Rodmann et al. 2006).
We summarize in Fig.~\ref{beta_obs} the results for the 
objects  with
7 mm   measurements from {\em Natta et al.} (2004) and references
therein and {\em  Rodmann et al.} (2006). 
For these objects,
the contribution from gas emission has been 
subtracted from the millimeter emission, and the spectral index
$\alpha$ has been computed  over the wavelength interval 1.3--7 mm;
in three cases the
7 mm flux is an upper limit,
once the gas emission has been subtracted from the total.
The values of $\beta$ have been derived
fitting disk models  to the data.
Note that these model-fitted $\beta$ values are slightly larger
than the optically thin determination $\beta=\alpha-2$,
typically by about 0.2. This difference is due to the fact that
the emission of the inner optically thick disk, albeit small,
is not entirely negligible.
The uncertainty on the model-derived 
$\beta$ is of about $\pm 0.2$ ({\em Natta et al.}, 2004;
{\em Rodmann et al.}, 2006). 
%In some case, there are independent measurements
%of $\beta$ by other authors, which agree with Fig.~\ref{beta_obs} estimates
%(GG Tau? HD 34282? others?).
The figure shows also the results for those objects for which no resolved maps exist, but have measurements of the integrated flux at 7 mm. For these, 
$\beta$ as been derived
from the measured $\alpha$ assuming that the disk is larger than $\sim$100 AU,
sufficient to make the optically thick contribution small.

\begin{figure}[ht]
 \epsscale{1.0}
%\plotone {ps.beta_PPV}
\plotone {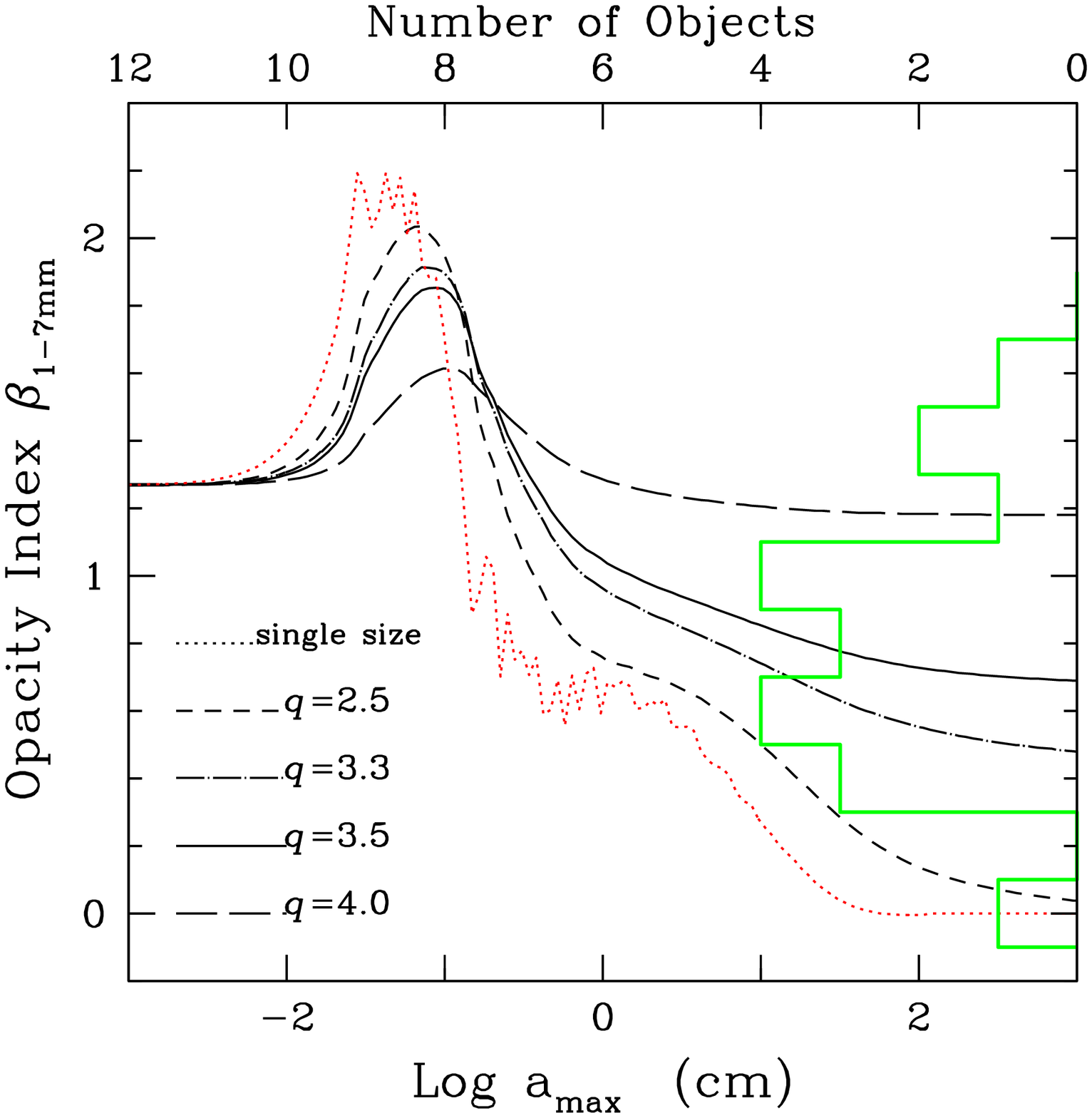}
\caption{ Opacity index $\beta$ for grains with a size
distribution $n(a)\propto a^{-q}$ between $a_{min}$ and
$a_{max}$. The index $\beta$ is computed between 1 and 7 mm for compact
segregated spheres of olivine, organic materials, water ice
({\em Pollack et al.}, 1994), and plotted as
function of $a_{max}$.  Different curves correspond
to different values of $q$, as labelled. In all cases, $a_{min} \ll 1$ mm.
The dotted curve shows $\beta$ for grains with single size $a_{max}$.
The histogram on the right shows the distribution of the 
values of $\beta$ derived from millimeter
observations (see Fig.~\ref{beta_obs}).  About 60\% of the objects have $\beta \le 1$, and 
$a_{max}\simgreat$ 1 cm. Different dust models give results which
are not very different (e.g., {\em Natta and Testi}, 2004).
}
\label {kappa_mm}
\end{figure}

If we consider only the 14 resolved disks,
which have all radii larger than 100 AU,
 $\beta$  ranges from 1.6 (i.e., very
similar to the ISM value), to 0.5. Ten objects
have $\beta\simless 1$, i.e., much flatter than ISM
grains. The unresolved disks  behave in a similar way;
the extreme case is
UX~Ori, with $\beta\sim 0$. Unfortunately,
the UX~Ori disk has not been resolved  
so far; at the distance of $\sim 450$ pc,
and, with an integrated flux at 7 mm of only 0.8 mJy ({\em Testi et al.}, 2001), it remains
a tantalizing object.
High and low  values of $\beta$  are found both for HAe stars and for TTS,
and there is no apparent correlation of $\beta$ with the stellar
luminosity or mass. Similarly, we do not observe any correlation
of $\beta$ with the age of the star. This last quantity, however, is often
very uncertain, and the observed sample limited.

%\bigskip
%\noindent
%\textbf{ HD~163296}
%Radial variation of $\alpha$

%\bigskip
%\noindent
%\textbf{ From $\beta$ to grain sizes}

Once we have established that the grains  originally in the collapsing cores,
for which $\beta \sim 1.7-2$, have gone through a large degree of processing,
one wants to derive from $\beta$ the properties of the actual grain
population, and in particular its size. 
The properties of the millimeter opacity of grains of increasing size have been
discussed by, e.g., {\em Beckwith et al.} (2000) and references therein.
Fig.~\ref{kappa_mm} illustrates the behaviour of $\beta$ computed between
1 and 7 mm for a population of grains with size distribution
$n(a)\propto a^{-q}$ between $a_{min}$ and
$a_{max}$. The index $\beta$ is plotted as function of $a_{max}$ for 
different values of $q$; in all cases, $a_{min} \ll 1$ mm.
One can see that, for all values of $q$,
as $a_{max}$ increases
$\beta$ is first constant at the value typical of grains of size $\ll 1$ mm,
it has a strong and rather broad peak at $a_{max} \sim 1$ mm
and decreases below the initial value for  $a_{max} \simgreat$ few mm. However,
only for $q <3$ $\beta$ goes to zero for large $a_{max}$; for $q>3$,
the small grains always contribute to the opacity, so that $\beta$ reaches
an asymptotic value that depend on $q$ and on the $\beta$ of the small
grains. For $q=4$, the asymptotic value is practically that of the
small grains.

The results shown in Fig.~\ref{kappa_mm} have been computed for 
compact segregated spheres of olivine, organic materials and water ice
({\em Pollack et al.}, 1994). 
The exact values of $\beta$ depend
on grain properties, such as their chemical
composition, geometrical structure and temperature
(e.g., {\em Henning et al.}, 1995); examples for different grain models
can be found, e.g., in {\em Miyake and Nakagawa} (1993),
{\em Kr\"ugel and Siebenmorgen} (1994), {\em  Calvet et al.} (2002),
{\em Natta et al.} (2004).
However,  these differences
do not undermine the general
conclusion that the observed spectral indices require that grains have grown
to sizes much larger than the observing wavelengths
(e.g., {\em Beckwith et al.}, 2000
and references therein, {\em Draine}, 2006).  To 
account for the observations which extend to 7 mm, objects with
$\beta\sim 0.5-0.6$ need a distribution of grain sizes with
maximum radii of few centimeter, at least.
The minimum grain radii are  not
constrained, and could be as small as in the ISM; however,
the largest
particles need to contribute significantly to the average
opacity, which implies that the grain size distribution cannot be
%steeper than $n(a)\propto a^{-(3\rightarrow3.5)}$;
steeper than $n(a)\propto a^{-(3\div3.5)}$;
in any case, the large grains  contain  most of the solid mass.
A detailed  analysis  can be found in 
e.g., {\em Natta and Testi} (2004).

Of the objects observed so far, TW~Hya is a particularly interesting case.
Millimeter observations can be fitted well with
models that include grain growth up to 1 cm ({\em Calvet et al.}, 2002;
{\em Natta and Testi}, 2004; {\em Qi et al.}, 2004). However, this is probably 
an underestimate of the maximum grain sizes in this disk.
{\em Wilner et al.} (2005) have demonstrated that the 3.5 cm emission of this
object, contrary to expectations, is not  dominated by gas emission, but
by thermal emission from dust grains in the disk, and that
a population of  particles as large as several centimeters
residing in the disk is needed to explain both the long wavelength SED and
the spatial distribution of the 3.5 cm emission.

\begin{figure}[ht]
 \epsscale{1.0}
%\plotone{fhd163_alp.eps}
\plotone {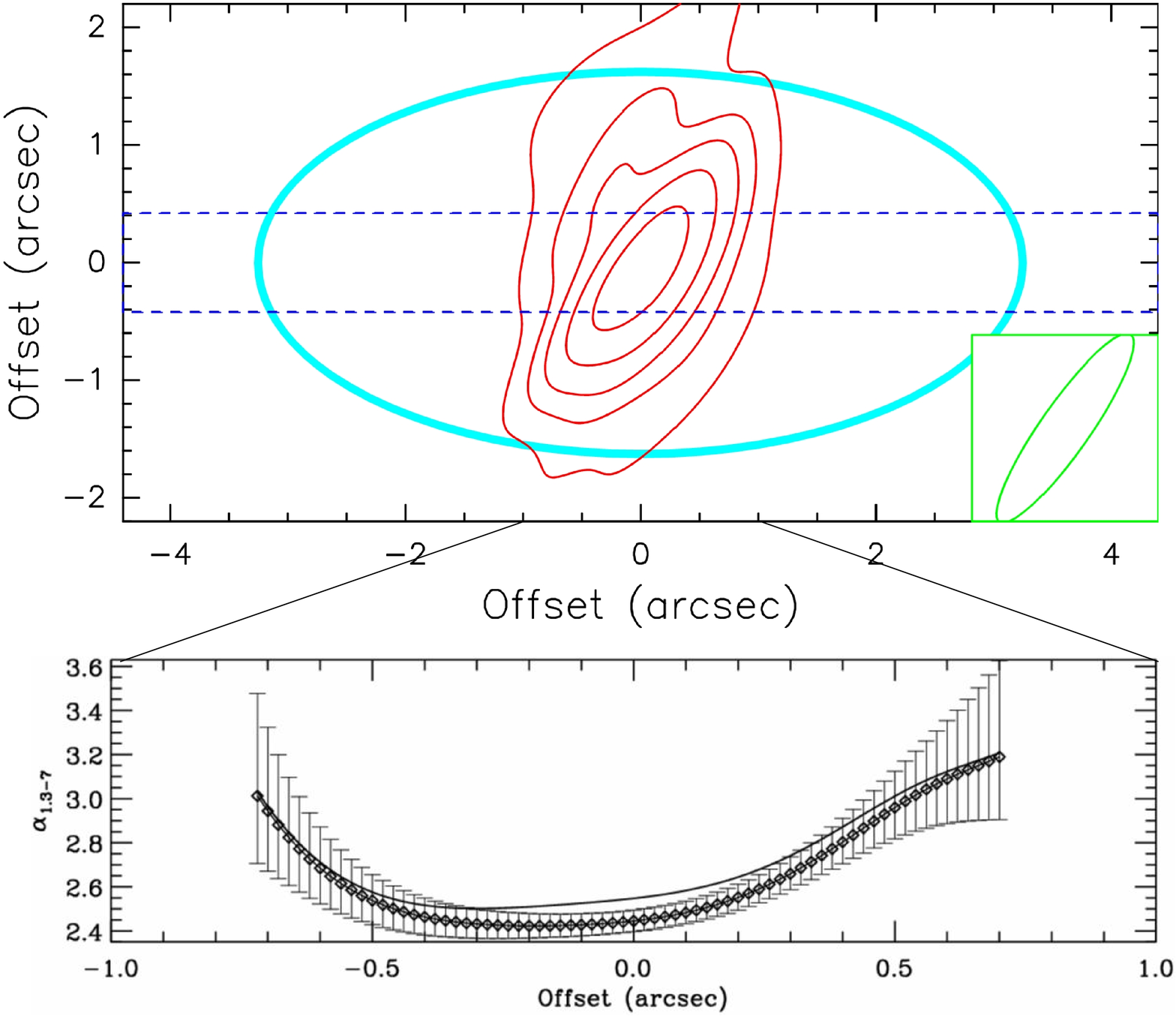}
\label{fhd163}
\caption{The HD163296 disk. In the top panel we show the IRAM-PdBI contour map
of the 1.3 mm emission from the disk, and the approximate shape and size 
of the scattered light disk as detected with HST/STIS.  The images have been rotated
so that the major axis of the disk is aligned with the abscissa of the plot. The 
spatial resolution of the millimeter observations is shown by the ellipse in the lower
right box of the upper panel. In the bottom panel the diamonds with error bars
show the variation along the disk major axis of the continuum spectral index
measured between 1.3 and 7 mm. In the same panel, the solid line shows the values
of the spectral index when a correction for the maximum possible gas contribution is 
applied to the 7 mm observations. The data suggest a  variation of the 
spectral index as a function of disk radius. If interpreted as a variation of the 
grain properties, they indicate that larger grains are found in the inner
($r\le 60$~AU) regions of the disk, while the grains in the outer disk may still
be in a less evolved stage.
}
\end{figure}

The result that dust particles have grown to considerable sizes, in fact to ``pebbles'',
is thus a solid one and verified in many systems. So far, it has been derived as
a ``global'' property of the dust grain population of the disk. In order to 
constrain dust growth models, the next logical step is to  check for 
 variation of dust properties as a function of radius. A first 
attempt at this type of study is shown in Fig.~\ref{fhd163}. High angular
resolution and high signal to noise 7 mm (VLA) and 1.3 mm (PdBI) maps 
of the HD~163296 system have been combined to reconstruct the millimeter spectral 
index profile as a function of radius. The result suggests that the spectral index varies
as a function of radius from $\alpha\sim  2.5$ in the inner $\sim$60~AU to $\alpha\sim 3.0$
in the far outer disk. If interpreted as differential grain growth,
this result implies that the outer disk contains less evolved particles than
the inner disk.

%{\bf Connection}
%The dust masses derived from the millimeter fluxes
%%, even using ``standard'' values
%%for the dust opacity coefficient (see, e.g., ??),
%are relatively large. These
%are, for most of the objects, lower limits to the real masses, as large grains
%are expected to have  opacities much smaller than the value we have used. 
%It is thus unlikely that a population of
%large bodies, to which our observations would not be sensitive, gives a significant
%contribution to the total disk solid mass. Of course, our observations would be 
%consistent with a scenario in which a small population of larger bodies have
%detached from the ``small'' grains ($\simless$ few cm)
%that contribute to the mm opacity.
%

%

\section{\textbf MINERALOGY}

The Infrared Space Observatory (ISO) and the latest generation
of mid-infrared spectrographs on large ground-based telescopes
have revealed the presence of many
new dust species, in addition to amorphous silicates,
in the protoplanetary disks surrounding 
pre-main-sequence stars. The position and strength of
the observed emission bands match well those of Mg-rich and Fe-poor crystalline
silicates of the olivine (Mg$_{2x}$Fe$_{2-2x}$SiO$_{4}$ and pyroxene
Mg$_{x}$Fe$_{1-x}$SiO$_3$ families, in particular their Mg-rich end
members forsterite and enstatite respectively (x=1).
Apart from these components, evidence for FeS and SiO$_2$
was also found ({\em Keller et al.}, 2002).
Spitzer is now providing similar results for stars of lower mass,
TTS and BDs.
%, as well as
%amorphous silicates, Polycyclic Aromatic Hydrocarbons (PAHs) and in one
%case aliphatic carbonaceous dust (emission at 6.9 $\mu$m in HD163296).
%For lower mass T~Tauri stars, ISOPHOT observations showed no evidence
%for crystalline silicates (Natta et  al. 2000).

The presence of these dust species in protoplanetary disks
implies substantial processing (chemical and physical) of the dust
because their abundance, relative to that of amorphous
silicates,  is far above limits set for these species
in the interstellar medium (e.g., {\em Kemper et al.}, 2004). Therefore, these dust species must be
formed sometime during the collapse of the molecular cloud core or
in the accretion disk surrounding the young star. There are
several possible mechanisms that may be responsible for grain
processing, and it is quite likely that several contribute. In the
innermost disk regions, where the dust temperature is above
1000~K,  heat will induce a change of the grain
lattice 
ordering (long-range ordering) from chaotic to regular 
(thermal annealing),
leading to the transformation of amorphous silicates into
crystalline ones (olivines and pyroxenes). Above about 1200-1300 K,
chemical equilibrium processes will lead to vapourisation and
gas-phase condensation of silicates, mostly in the form of
crystalline forsterite. Local high-energy processes (shocks, lightning) as
well as radial mixing may increase substantially the  amount of crystalline
silicates in the cool outer regions of the disk. Note that thermal
annealing of amorphous, presumably Fe-rich silicates would lead to
Fe-rich crystalline silicates. Therefore the Fe content of
crystalline silicates may be used as an indication of the chemical
processing that has occurred (gas-phase condensation, annealing,
local processes). In this context, we  stress the
importance of improving the
still poorly constrained stochiometry of  interstellar amorphous
silicates, in particular the Fe/Mg and Mg/Si ratio.

The mineral composition of dust in protoplanetary disks can be
compared directly to that of solar system comets, asteroids and
interplanetary dust particles (IDPs). Only one high-quality 2-200
 $\mu$m spectrum of a solar system comet, Hale-Bopp, is available from
ISO. The crystalline silicates in Hale-Bopp are very Mg-rich and
Fe-poor, similar to those in protoplanetary disks. Laboratory studies
of IDPs of cometary origin also reveal the presence of Mg-rich
crystalline silicates. The situation for asteroids is much more
diverse. Apart from Mg-rich crystalline silicates, many asteroids show
Fe-containing silicates, probably related to parent body processing or
to nebular processing in the inner solar nebula. 
%\emph{Spitzer comet results?}

\bigskip
\noindent
\textbf{7.1
Spatially unresolved 10 $\mu$m spectra}

Information on the abundance of crystalline silicates 
in a large sample of objects of different mass are derived
from the shape of the 10 $\mu$m feature (mostly
through the 11.3 $\mu$m
forsterite peak) in spatially
unresolved observations.
Fig.~\ref{sil_prof}  shows how this feature can be prominent in stars
of all mass, from HAeBe to brown dwarfs.
Fig.~\ref{sil_ratios}
%, which  provides a simple characterization of the 10 $\mu$m profiles, 
shows how in many objects the flux at 11.3 $\mu$m is stronger than
at 9.8 $\mu$m, where the emission of small amorphous
silicates roughly peaks. A strong 11.3 $\mu$m peak
is typical of profiles with a
large  component of crystalline silicates, in addition to
the amorphous ones. This interpretation has been confirmed by
the presence of crystalline features at longer wavelengths, when
the data are available.   Contamination from
the 11.3 $\mu$m PAH feature has been ruled out in many cases,
from the absence  of the other PAH features at shorter wavelengths
(see, for example, {\em Acke and van der Ancker}, 2004). 

A determination of the fractional mass abundance of the crystalline
silicates has been derived by various authors  fitting the
observed profiles with mixtures of grains of different
composition (typically, olivines and pyroxenes), size (from submicron
to few microns)
and allotropic state (amorphous and crystalline). In general,
the models assume a single temperature for all grains, and that
the emission is optically thin. These assumptions, and the limited
number of dust components that are included in  models, 
make the estimates very uncertain. However, the results are interesting. 
%as shown in Fig.~\ref{sil_crist}.
For HAeBe stars, {\em van Boekel et al.} (2005) find
that crystalline silicates 
are abundant only when the amorphous silicate grain
population is dominated by large ($\geq$ 1.5 $\mu$m) grains. In
Fig.~\ref{sil_ratios}, there are no points with 11.3/9.8 ratio $\simgreat 1$
and large values of $F_{peak}/F_{cont}$. The fraction
of grains which is crystalline ranges from 5\% (detection limit)
to $\approx$ 30\%.  Note that grains larger than few
microns are not accounted for; if crystalline silicates are mostly
sub-micron in size, while the amorphous species have a much broader
distribution, the degree of crystallinity can be much lower
than current estimates. This seems to be the case in
Hale-Bopp ({\em Min et al.}, 2005), and may happen in protostellar disks
as well.
In addition,  the derived crystalline/amorphous abundances are spatially
averaged values; there are likely strong radial gradients in
crystallinity (see below). 
As more data for lower mass objects become available,
we see
that they show a similar trend, although   relatively
high crystallinity and relatively few large amorphous grains
have been seen in one very low mass TTS 
(e.g., {\em Sargent et al.}, 2005).
Recent Spitzer data for brown dwarfs in ChaI show that 
also objects of very low mass
have large fraction of crystalline silicates, from 9 to about 50\%
({\em Apai et al.}, 2005).

%There is no clear correlation between star mass and disk crystallinity.
%Higher values in BDs? possible interpretation.

%T~Tau stars in general show less evidence
%for crystalline silicates (Przygodda et al. 2003), but there are
%notable exceptions (Honda et al. 2003).

As far as the chemical composition of the silicates is concerned, there are
some general trends, showing that crystalline pyroxene grains can be
found in disks with relatively large fractions of forsterite
and silica, but they are typically not
found in disks with small fractions of forsterite and silica
({\em van Boekel et al.}, 2005; {\em Sargent et al.}, 2006).

%\begin{figure}[ht]
% \epsscale{1.0}
%\plotone{ps.silcrist_mstar}
%\caption{Mass fraction of crystalline silicates as function
%of the stellar mass for  BDs  ({\em Apai et al.}, 2005; stars),  
%TTS ({\em Bouwman et al.}, 2005; {\em Sargent et al.}, 2005;
%{\em Przygodda et al.}, 2004; squares), 
%HAeBe {\em van Boekel et al.}, 2005; filled dots). 
%Note that the values for different samples may suffer from systematic
%differences in the models used for fitting; this is particularly
%true when comparing the  {\em Sargent et al.} sample to the others.
%}
%\label {sil_crist}
%\end{figure}

\bigskip
\noindent
\textbf{7.2
Spatially resolved spectroscopy}

\begin{figure}[ht]
 \epsscale{1.0}
%\plotone {MIDI_plot.eps}
\plotone {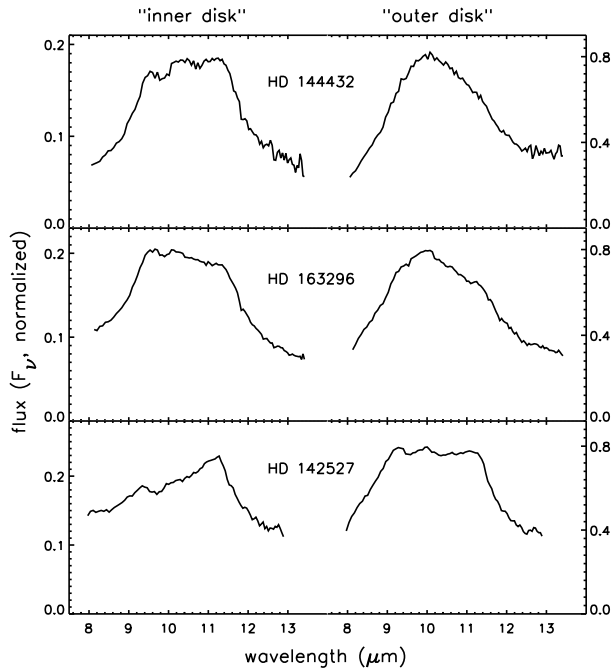}
\caption{Comparison of the silicate profiles from the inner
(1--2 AU) and outer (2--20 AU) disk in three HAe stars
({\em van Boekel et al.}, 2004a). The data have been obtained with the
mid-IR interferometric instrument MIDI on the VLTI. The comparison
shows that, for all the three stars, the crystalline features are much stronger
in the inner than in the outer disk.}
\label {MIDI}
\end{figure}

The advent of spectrally resolved mid-infrared interferometry using large baselines at
the Very Large Telescope Interferometer (VLTI) 
%and the Keck Interferometer (KI) 
has made it possible for the first time to study
the nature of the dust grains on AU spatial scales. The MIDI
instrument at the VLTI has been used to study the innermost regions of
HAeBe star disks. 
%these are big and bright and thus ideal sources
%for first science with VLTI. 
van Boekel et al. (2004a; see Fig.~\ref{MIDI})
show that the inner 1-2~AU of the disk surface surrounding three HAe
stars is highly crystalline, with between 50 and 100 per cent of the
small silicates being crystalline. However, crystalline silicates
are also present at larger distance, with
a wide range of abundances.  A similar behaviour is found in
other objects observed with MIDI ({\em Leinert et al.},
2006).
Clearly, large star-to-star variations
in the amount and distribution of the crystalline silicates
exist, but it seems that crystalline grains are relatively more
prominent in the inner than in the outer disk.
Note that, because silicate emission from cold grains become
very weak, in this case "outer" refers to warm regions of the disk
between 2 and 20 AU
at most ({\em van Boekel et al.}, 2004a).

In the star with the highest crystallinity, HD~142527, the
observations suggest a decrease in the ratio between forsterite and
enstatite with increasing distance from the star. Such a trend in the
nature of the crystals with distance is predicted by chemical
equilibrium and radial mixing models ({\em Gail}, 2004). The innermost disk
is expected to contain mostly forsterite, while at larger distance
from the star a conversion from forsterite to enstatite takes
place. Thermal annealing produces both olivines and pyroxenes, with a
ratio depending on the stochiometry of the amorphous material accreted
from the parent molecular cloud.  {\em van Boekel et al.} (2004a) also find
evidence for a decrease of the average grain size with distance from
the star, suggesting that grain aggregation has proceeded further in
the innermost, densest disk regions. 
%Such a trend is expected from
%grain growth and settling models (reference Carsten?).

\bigskip
\noindent
\textbf{7.3
TW Hya Systems}

A number of objects with distinct signs of  evolution
of their inner disk has been identified in the last decade.
These objects with ``transitional'' disks are characterized by
spectral energy distributions (SEDs) with
a flux deficit in the near-infrared relative to the 
median SED of  Classical T Tauri stars in Taurus (a proxy
for the expected emission from optically thick disks, 
see {\em D'Alessio et al.}, 1999),
while at longer wavelengths fluxes are comparable or sometimes
higher than the median SED of Taurus. Ground-based near and mid-infrared
photometry combined with IRAS fluxes allowed the identification
of objects with these properties in Taurus ({\em Marsh and Mahoney}, 1992;
{\em Jensen and Mathieu}, 1997;
{\em Bergin et al.}, 2004) and in the TW Hya association 
({\em Jayawardhana et al.}, 1999; 
{\em Calvet et al.}, 2002).
However, instruments on board Spitzer are providing a much better view
of the characteristics and frequencies of these objects ({\em Uchida et al.}, 2004;
{\em Forrest et al.}, 2004;
{\em Muzerolle et al.}, 2004; 
{\em Calvet et al.}, 2005; 
{\em Sicilia-Aguilar et al.}, 2006). 
The inner disk clearing may be related to photo-evaporation of the outer
disk by UV radiation ({\em Clarke et al.}, 2001). However, transitional disks
have now been found in brown dwarfs ({\em Muzerolle et al.}, 2006), for which
the UV supply of energy is negligible. An alternative explanation is that
a planet has formed and  carved out a gap in the disk; hydrodynamical
simulations seem to support this view ({\em Rice et al.}, 2003;
{\em Quillen et al.}, 2004).

The 10 Myr old TW Hya was the first of these objects analyzed in detail.
The peculiar SED of this object ({\em Calvet et al.}, 2002;
{\em Uchida et al.}, 2004),
can be understood by an optically thick disk truncated at $\sim$ 4 AU;
the wall at the edge of this disk is illuminated directly by stellar
radiation, producing the fast rise of emission at wavelengths $> 10 \mu$m.
The region encircled by this wall is not empty. For one thing, it has
gas, because the disk is still accreting mass onto the star ({\em Muzerolle et al.}, 2000).
In addition, this inner region
 contains $\sim$ 0.5 lunar masses of micron-size particles,
responsible for the small near-infrared flux excess over photospheric emission. 
This dust is also 
responsible for the strong silicate feature at 10 $\mu$m, which
allows us to 
examine the conditions of the dust in the inner disk ({\em Calvet et al.}, 2002).
The profile of the 10 $\mu$m feature, integrated over the disk, is
typical of amorphous silicates with sizes of $\sim 2$ $\mu$m
({\em Uchida et al.}, 2004), with a very low content of crystalline silicates,
$< 2$ \% fraction by mass 
({\em Sargent et al.}, 2005).
However, interferometric
observations with
MIDI  resolve the feature from the inner disk (within
$\sim$1 AU; see {\em Leinert et al.} 2006 and the chapter by
{\em Millan-Gabet et al.}),
showing clearly the 11.3 $\mu$m peak typical of forsterite, while
the uncorrelated profile
 is very similar to that measured with single-dish instruments.  

{\em Sargent et al.} (2006)
 have analyzed the dust content of a number of transitional
disks, in addition to TW Hya. They find, averaged
over the whole disk, a very low content of crystalline silicates.
In contrast, there is
a large spread of the crystalline silicate fraction 
(from 0.3\% to 20\% in mass)
in their sample of objects with optically thick inner disks.

%This lack of silicate crystallinity is consistent with the absence of
%crystalline silicates in the interstellar medium (Kemper et al. 2004),
%suggesting both that silicate material in the outer regions of disks
%of T Tauri stars originates in the ISM and that the silicates in these
%outer regions are little processed.
These results support the idea that
disks accumulate
crystalline silicate grains in their innermost regions 
%either by inner disk
%heating  or by local annealing of silicate grains, 
and that  radial mixing 
transports outward some fraction of this
material, with an efficiency that decreases with radius. 
In transitional disks, a large fraction of
the  region within $\sim$few AU  is cleared out, leaving
only material  with low 
crystalline content. Spatially resolved observations of the
innermost regions are needed to detect the small amount of remaining grains,
with their higher crystallinity.

%\section {PUTTING THINGS TOGETHER}
%
%As discussed in Sec.3, the information we have on grains in disks is by
%necessity fragmentary. We have no access to the properties of grains
%in the disk midplane, but for regions where the disk becomes
%optically thin, typically at $R\simgreat 10-30$ AU. In
%this outer disk, however, grains are too cold to emit at 10 and 20 \um,
%so that we have only information on the .. Inside this
%radius, we  only know the properties of grains on a thin surface layer,
%which contains only a tiny fraction of the disk mass.
%The most used diagnostic of grain properties in this region, the
%emission feature of silicates, can only sample warm dust, so that
%we have no information on the presence and fraction of silicates
%smaller than few microns in the outer disk, nor of their mineralogy.
%
%Nevertheless, it is interesting to see what happens in objects for which we
%have the largest set of data. 
%
%(some van den Ancker paper?).
%
%We have collected the information on the properties of the silicate feature
%for objects with well measured millimeter properties (Table 1).
%
%Butterfly nebula?
%....
\section{\textbf SUMMARY AND OPEN PROBLEMS}

\bigskip
\noindent
\textbf{8.1 Evidence of grain growth}

The  results we have discussed  come from a variety of techniques, including
optical,   near and mid-infrared imaging,
mid-infrared spectrometry and millimeter interferometry.
All the data  show  clear evidence that grains
in protoplanetary disks  differ
significantly from grains in the diffuse ISM and in
molecular clouds. They are  much  larger on average,
and there is increasing support for a scenario in which grains grow and
sediment toward the disk midplane. 
%We should, however, remember that the observations give us
%a piecemeal picture only. 
On the surface of the disk, within few tens AU from the
star, we know that in most cases silicates have
grown to sizes of few microns, although much smaller particles
(PAHs) may be present as well, and there are hints of vertical
stratification (at several scale heigths from the midplane)
of these micron-size grains from studies of
scattered light and silhouette disks.
We have evidence from 
millimeter interferometry that, if we consider not
only grains on the disk surface but the bulk of the dust mass,
in many disks  grain growth has not stopped
at  micron-sizes,  but there is
a dominant
population  of  ``sand and pebbles'', millimeter and
centimeter grains.
% which account for a large dust mass, probably
%close to the total solid mass in the disk.
In addition to vertical gradients of grain sizes, there is also
evidence that radial gradients of grain properties
exist, as shown by mid-IR interferometry,
absorption mid-IR spectra of silhouette disks,
and millimeter images at different wavelengths, as
discussed before. Grains in the very outer disk seem to be less
processed than grains closer to the star.

%This picture of grain growth, however, is difficult to be defined more quantitatively.
One important  aspect to keep in mind is that
not all the objects show evidence of  processed 
grains: in some stars, for example, 
the 10 \um\ silicate  feature has a shape very similar
to that of the small silicates in the ISM; some disks have a rather steep
dependence of the millimeter flux on wavelength, again typical of the
small ISM grains. Some objects seem to have small
silicates on the disk surface, while very large grains are
implied  by the millimeter spectral shape.
An impressive case is that of UX~Ori, which has
a very peaked
and strong 10 \um\ emission feature, typical of the small ISM amorphous
silicates (Fig.~\ref{sil_prof}) and  
an extremely flat millimeter spectrum, which may
indicate the presence of very large grains in the midplane.

One should also
remember that the very nature of protoplanetary disks
(i.e., high optical depth, strong temperature radial gradient, etc.)
prevent a complete census
of the dust population, so that 
%we cannot determine,
%e.g.,  the size distribution of the
%various materials over the whole range of grain dimensions we know to exist.
deriving from the observations
the global properties of the solids in any given disk
(such as  the  density of grains of different size and composition
as function of radius and altitude) remains for the moment
impossible, and we have to do with a piecemeal picture.

\bigskip
\noindent
\textbf{8.2 Dependence on stellar properties}

One important step forward since PPIV has been the capability to study
disks around stars of very different properties.
The advent of 10-meter class telescopes on the ground, the success of
Spitzer and the improvement of the millimeter interferometers, 
including the VLA, have given us access to objects of increasingly
lower mass. We have now millimeter spatially resolved data for TTS,
and mid-IR spectra for very low mass stars and brown dwarfs, which
allow us to investigate if there is a dependence of grain evolution
on stellar and disk parameters.
It may be worth noting that,
within the millimeter-observed sample, which includes TTS and HAe stars, 
the stellar mass varies by a factor 5 and the
luminosity by a factor
250; assuming typical scaling laws ({\em Muzerolle et al.} 2003;
{\em Natta et al.}, 2004b, 2006), the accretion rate in the disk
is likely to vary by at least a factor 25. 
%the X-ray luminosity by a factor ??;
The range of stellar properties are
even larger for the mid-infrared sample, which includes
now several brown dwarfs and covers the  interval  between
$\sim 0.04$ and 2--3 \Msun\ in mass,  $\sim 0.01$
and 50 \Lsun\ in luminosity, $\sim 10^{-10}$ and
$10^{-6}$ \Msun/yr in accretion rate.  Over this large range of physical
conditions,
we do not detect {\em systematic} differences, i.e.,   grain
properties do not seem to correlate with
any star or disk parameter. 
At any given mass, there are large variations of grain properties,
although objects with unprocessed grains seem to be rare.

It seems that disks around all kind of stars
have the potential of processing
 grains,
to a  degree that varies for reasons  not understood so far.
%The interpretation of the observations can be complicated also
%by the existence (predicted theoretically and now observed in few cases)
%of radial gradients in grain properties. The ``transitional disks",
%where the clearing of the inner region results in a very low fraction
%of crystalline silicates in spatially unresolved spectra, is a case in point.
%Spatially resolved observations will be able to answer this point.
%Also, some help can come from those (few) objects where grains have
%properties similar to the ISM ones, i.e., where there has been very
%little processing. 

\bigskip
\noindent
\textbf{8.3 Time dependence}

The  time dependence of grain growth is a crucial information
for models, and one that we would like to derive from the
observations with how much detail as possible.
We should stress, first of all,  that
there is not, at present, anything close to an {\em unbiased} sample
which can be used to study grain evolution. All the observing techniques
discussed in this review have been applied to selected objects, known
to be suitable for detection and analysis.  
%This is, unfortunately, still very difficult.
In addition,
ages of individual pre-main sequence stars (our best clock  so far)
are measured from the
location on the HR diagram and have large uncertainties, especially
for HAe stars and  brown
dwarfs. Even more importantly, the objects for which we have
observations vary in age 
by about a factor of ten only (from
$\sim 1$ to $\sim 10$ Myr), and individual scatter may dominate
over an underlying time dependence.

The data available so far show no evidence of a correlation
of the grain properties (at any wavelength) with time:
we  find very processed grains in the disks of the youngest as 
of the
oldest pre-main sequence stars.  
One possible explanation is 
that grains are processed efficiently
in the very early stages of disk evolution, when the system star+disk
is still embedded and  accreting actively from the natal core
(see, e.g., the theoretical models of dust dynamics and evolution
during the  formation of a protostellar disk by 
{\em Suttner and Yorke}, 2001). 
Then, the large star-to-star variation seen
in, e.g., the HAe stars would be 
 due to differences in processing during
this initial, very active phase.
However, we cannot exclude that
further modifications also
take place in the $\sim 10$ Myr in which the disk continues to
exist.

Some information can  be obtained by studying grain properties in disks in
Class I sources,
which are  in an earlier evolutionary stage than the
disks around optically visible objects discussed in this review.
In Class I objects,
the silicate features are seen in general in absorption, and show
profiles typical of small,  unprocessed amorphous silicates.
It is likely, however, that the absorption is dominated by grains in
the surrounding envelope, for which we expect very little growth.
Only if the balance between the disk emission and the intervening absorption
in the core is favorable, can we have a glimpse at the properties of the
disk silicates, and in few cases there is evidence that the disk
silicates are much larger than those in the core
(e.g., {\em Kessler-Silacci et al.} 2005).
Also, 
the young protostellar binary SVS20 in Serpens shows evidence of the 11.3 \um\
forsterite feature in its spectrum ({\em Ciardi et al.}, 2005).
For the moment, there are only a few good quality spectra,  but
we can expect major progress soon, as Spitzer data will become available.
One would also like to know if Class I disks have  millimeter
spectral energy distributions as flat  as those of Class II objects.
Unfortunately,
using millimeter interferometry to measure the spectral index of the
emission of disks embedded in Class I cores has proved so far very
difficult, as the core emission is substantial even at the smallest
physical scales one can probe.
The extremely flat spectra of the few objects studied so far come
from unresolved central condensations, and cannot be interpreted
as evidence of 
very large grains in disks 
until the emission can be spatially resolved ({\em Hogerheijde et al.}, 1998,
1999).

\bigskip
\noindent
\textbf{8.4 Comparison with grain evolution models}

Models of grain growth by collisional coagulation and sedimentation
(e.g., {\em Dullemond and Dominik}, 2005; {\em Weidenschilling and Cuzzi}, 1993,
and references therein) predict that grain growth will occur
on very short time scales. 
This, at first sight, is in agreement with the idea that grain properties
change already in  the  early  evolutionary phases, when disks are
actively accreting matter from the parental core.

However, models also predict that the growth will not stop at micron or
even at centimeter sizes, but will proceed very quickly (compared to
the pre-main sequence stellar life times) to form
planetesimals. This is not consistent with the observations, which
show us disks around stars as old as $10^7$ years with 
grains that, albeit very large, are  far from being planetesimals.
When fragmentation of the conglomerates is included, in many cases
the outcome is a bimodal distribution for the solids, with most
of the mass in planetesimals and a tail of smaller fragments, which can be as large as centimeters and meters, but contains only a tiny
fraction of the mass in planetesimals.
If these smaller bodies provide sufficient optical depth, the disk
properties may  look similar to what is observed.
However, we think this is unlikely to be the case.
The millimeter observations
described in Section~6 tell us not only that grains need to have very
big sizes but also the mass of these grains, which turns out to be
very large. Using, e.g., the same grain models as in Fig.~\ref{kappa_mm},
we derive  dust  masses, for the disks with evidence
of very large grains,  between $10^{-3}$ and $10^{-2}$
\Msun;
assuming the standard gas-to-dust ratio of 100 (which is probably close
to true when disks formed), this implies disk masses between
$\sim$ 0.1 and 1 \Msun, i.e., ratios of the disk to star masses 
approaching unity.  If a large fraction of the
original dust is not in the observed population of millimeter and centimeter
grains, but in planetesimals, the disk mass should be even higher,
 well above the gravitational instability limit.
Although we cannot exclude that a {\em small} fraction of the
original solid mass is in planetesimals, we think it is unlikely
that they have collected  most of it. 

The
survival of a large mass of millimeter and centimeter-size grains
over a time scale of several million years, as observed
in many disks,  is challenging the models on several grounds.
It suggests that in
many cases the process of planetesimal formation
is  much less efficient than
predicted by theoretical models.
If disks are forming planetesimals as fast as predicted,
then this can only happen 
at the end of a long quiescent phase, where growth
is limited to sizes of few centimeter at most; otherwise, the process
has to be  very slow, involving a modest fraction of the dust mass
for most of the pre-main sequence life of the star.
The ``inefficiency" may   result from  fragmentation of the larger
bodies. This in-situ production of the grains we observe may 
take care of the other severe difficulty one has, namely that
millimeter and centimeter solid bodies (in a gas-rich disk)
migrate toward the star on very
short timescales.  This is a very open field, where progress
can be made only by combining together theoretical studies,
laboratory experiments and observations.

It is  possible that
the disks we can study with current techniques  are those
that will never form planets.
% but will dissipate on longer timescales
%or different reasons, e.g.,  viscous evolution,
%hotoevaporation, tidal interactions with
%companions etc. 
Disks where planets form may
evolve indeed very quickly, and could be below
present day detection limits.
This, however, seems unlikely, as we have 
objects like TW~Hya and HD~100546, which have evidence of very large grains
in their outer disk, and an  inner gap possibly due to the action of
a large planet.  Similar ``transitional" objects are found by Spitzer in 
increasing number, and a detailed characterization of their millimeter
properties will be important.
In any case, the  properties and evolution of disks that
we have studied so far and their dust content
provide the only observational constraints to theoretical models of grain evolution
and planet formation.

\bigskip
\newpage
\noindent
\textbf{8.5 Crystalline silicates and mineralogy}

The ISO discovery of large fractional amounts of crystalline silicates in disks
around stars of all masses has been a surprise.
They are  enhanced in the innermost regions of disks, as shown by
interferometric observations of HAe stars and by
the spectra of transitional disks, but they are also present, in variable
degree, further out.
Fractional masses of crystalline silicates, integrated over
the whole disk emitting regions,
have been derived for many objects,
HAe, TTS and BDs, with values
ranging from zero to about 60\%.  These  estimates
refer to grains  on the disk surface, typically within 10--20 AU from the star
for the HAe stars, and 0.1--0.2 AU for BDs.
One should note that these values are extremely model dependent, 
and refer to a limited range of grain sizes only;
they should be taken with the greatest care.
We can expect   great progress in the immediate future,
as spectra in
the 15-45 $\mu$m spectral
region, which can detect  colder (and larger) crystals, are becoming
available. Further insight can also
come from improved characterization of the mineralogy in disks,
e.g., measuring the ratio of species such as crystalline pyroxene, 
forsterite, silica etc.
Self-consistent  radiation transfer models in disks which 
can include
a large number of dust species exist, and can be used to provide  
more reliable values for  the relative abundances of the various
species observed on disk surfaces.

The formation of  crystalline silicates requires energetic 
processes, which have to be efficient 
not only in HAe stars, where crystalline silicates were first
discovered, but also in the much less luminous TTS and brown dwarfs.
If crystallization is restricted to the very inner disks, as in
the {\em Gail} (2004) models, then  strong radial mixing must
occur, transporting material outwordly.
Radial and meridian drifts can have an important role in 
the growth of grains, which need to be addressed in the future.

\bigskip
\noindent
\textbf{8.6 Final remarks}

The observations we have discussed in this chapter show clearly
that very few disks (if any), contain unprocessed grains, i.e., with
properties similar to those of grains in the ISM.
In general, dust has been largely processed in all objects.

However, the degree and the result of these changes vary 
hugely from object to object, as we find large differences
in the grain size, composition and  allotropic properties.
At present, we have not been able to identify any correlation
between the grains and other properties of the central star
(such as its mass and luminosity) and  of the disk (for example,  mass,
accretion rate, etc.), nor with the age of the system.
This last point is particularly distressing, since 
planet formation theories require
to understand if, when and how grains change with time,
or, in other words, how they ``evolve".

The study of pre-main sequence stars has shown us that, no matter
which aspect of their rich phenomenology we are interested in,
there is a large scatter between individuals, which can hide
underlying trends. 
%A recent example is the correlation between
%the mass accretion rate and the mass of the central object,
%which has become clear only when brown dwarfs and intermediate-mass
%stars have been added to TTS, expanding the range of masses to
%about two order of magnitudes. 
In discussing grain properties,
it is possible that the objects observed so far do not sample the right
range of parameters, age in particular, or that 
the available samples are still too small, or
restricted to a too narrow range in the parameter space.
We see the noise, and cannot
identify trends which remain, if present, hidden.
If this is really the case, we can expect major advances in the near future,
as new space and ground-based facilities  will make possible to
study much larger  samples of disks, in different star-forming
regions, surrounding objects  distributed over a
broader range of mass,  age and  multiplicity.

\vskip 0.3cm

\textbf{ Acknowledgments.}  We are indebted to a number of collegues, who
have provided to us unpublished material and helped in
preparing some of the figures. Among them,
Jeroen Bouwman, Roy van Boekel, Ilaria Pascucci,
Daniel Apai, Jackie Kessler-Silacci, Jens Rodmann, Michiel Min,
and the IRS Spitzer team. During a visit to the MPA in Heidelberg, A.N. 
enjoyed   discussions with Kees Dullemond, Christoph Leinert,
Jeroen Bouwman, Roy van Boekel and Jens Rodmann, among others.
This work was partially supported by MIUR grant 2004025227/2004
to the Arcetri Observatory.

\bigskip
%\newpage

\centerline\textbf{ REFERENCES}
\bigskip
\parskip=0pt
{\small \baselineskip=11pt

\refs Acke B. and van den Ancker M. E. (2004) \aap, {\em 426}, 151-170.

\refs Apai D., Pascucci I., Bouwman J., Natta A., Henning Th., and
Dullemond C. P. (2005) {\em Science}, {\em 310}, 834-836.

\refs Beckwith S. V. W. and Sargent A. I. (1991) \apj, {\em 381}, 250-258.

\refs Beckwith S. V. W., Henning Th., and Nakagawa Y. (2000).
%Dust properties and assembly of large particles in protoplanetary disks. 
In {\em Protostars and Planets IV} (V. Mannings, A. P.
Boss, and S. S. Russell, eds.), Univ. of Arizona Press, Tucson,
533-558.

\refs Bergin E., Calvet N., Sitko M. L., Abgrall H., D'Alessio P., 
et al.
%Herczeg G.J., Roueff E., Qi C., Lynch D.K., Russell R.W., and 2 coauthors 
(2004) \apj, {\em 614},  L133-L136.

\refs Bianchi S., Gon\c{c}alves J., Albrecht M., Caselli P., Chini R., Galli D., and Walmsley M. (2003) \aap, {\em 399}, L43-L46.

\refs Blum J. (2004) 
%Grain growth and coagulation. 
In {\em Astrophysics of Dust} (A.N. Witt, G.C. Clayton, and B.T. Draine,
eds.), pp. 369-391. ASP, San Francisco.

\refs Bouwman J., Henning Th., Hillenbrand L., Silverstone M., 
Meyer M., Carpenter J., Pascucci I., Wolf I., Hines D., and Kim J.
(2006), in preparation.

\refs Calvet N., Patino A., Magris G.C., and D'Alessio P.
(1991) \apj, {\em 380}, 617-630.
% Median of TTS SEDs

\refs Calvet N., D'Alessio P., Hartmann L.,
         Wilner D., Walsh A., and Sitko M.  (2002) \apj, {\em 568}, 1008-1016.
%TWHya

\refs Calvet N.,  D'Alessio P., Watson D. M., Franco-Hern\'andez R., 
Furlan E., et al.
%Green J., Sutter P. M., Forrest W. J., Hartmann L., Uchida K.I., and 6 coauthors 
(2005) \apj, {\em 630}, L185-188. 
%2 transitional TTS with Spitzer

\refs Chiang E. I. and Goldreich P. (1997) \apj, {\em 490}, 368-376.

\refs Chiang E. I. and Goldreich P. (1999) \apj, {\em 519}, 279-284.

\refs Chokshi A., Tielens A. G. G. M., and Hollenbach D. (1993)
\apj, {\em 407}, 806-819.

\refs Ciardi D. R., Telesco C. M., Packham C., G\'omez Martin C.,
 Radomski J. T., De Buizer J. M., Phillips Ch. J., and Harker D. E. 
(2005) \apj, {\em 629}, 897-902.

\refs Clarke C. J., Gendrin A., and Sotomayor M. (2001) \mnras, 
{\em 328}, 485-491.

\refs D'Alessio P., Calvet N., Hartmann L., Lizano S., and Cant\'o J.
(1999) \apj, {\em 527}, 893-909.
% median sed of TTS

\refs D'Alessio P., Calvet N., and Hartmann L. (2001) \apj, {\em 553}, 321-334.
%Grain growth, contiene HK Tau/C

%\refs D'Alessio, P., et
%al.\ 2005a, \apj, 621, 461

%\refs D'Alessio P., Calvet N., Hartmann L., Franco-Hernandez R.,
%\& Servin H. 2005b, ApJ, submitted

\refs Draine B. T. (2003) {\em Ann. Rev. Astron. Astrophys., 41}, 241-289.

\refs Draine B. T. (2006) \apj, {\em  636}, 1114-1120.

\refs Duch\^{e}ne G., McCabe C., Ghez A. M., and Macintosh B. A.
(2004) 
%A multiwavelength scattered light analysis of the dust
%grain population in the GG Tauri circumbinary ring. 
{\em Astrophys. J., 606}, 969-982.

\refs Dullemond C. P. and Dominik C. (2005) {\em Astron.
Astrophys., 434}, 971-986.

%\refs Fabian, D., J\"ager, C., Henning, Th., Dorschner, J., and
%Mutschke, H. (2000). 
%%Steps toward interstellar silicate mineralogy. V. Thermal evolution of amorphous magnesium silicates and silica. 
%{\em Astron. Astrophys., 364}, 282-292.

%\refs Fischer, O., Henning, Th., and Yorke, H.W. (1996).
%%Simulation of polarization maps. II. The circumstellar environment
%of pre-main sequence objects. 
%{\em Astron. Astrophys., 308}, 863-885.

\refs Forrest W. J.,  Sargent B., Furlan E., D'Alessio P., Calvet N., 
et al.
%Hartmann L., Uchida K. I., Green J. D., Watson D. M., Chen C. H., 
%and 11 coauthors 
(2004) \apjs, {\em 154}, 443-447.

\refs Gail H.-P. (2004) \aap, {\em 413}, 571-591.

\refs  Guillois O., Ledoux G., and  Reynaud C. (1999) \apj, {\em 521}, 
L133-L136.

\refs Habart E., Natta A., and Kr\"ugel E. (2004a) \aap, {\em 427}, 179-192.

\refs Habart, E., Testi, L., Natta, A., and Carbillet, M. (2004b) \apj, {\em 214}, L129-132.

\refs Habart E., Natta A., Testi L., and Carbillet M. (2006) \aap, 
in press (astro-ph 0503105)

\refs Henning Th., Michel B., and Stognienko R. (1995).
  {\em Planet. Space Sci., 43}, 1333-1343. 
  
%\refs Henning, Th.(ed.). 2003. {\em Astromineralogy}, Springer-Verlag,
%Berlin.  
  
\refs Henning Th., Dullemond C. P., Wolf S., and Dominik C.
(2005)
% Dust coagulation in protoplanetary disks, 
In {\em Planet
Formation. Theory, Observation and Experiments}, (H. Klahr, W.
Brandner, eds.), in press. Cambridge University Press, Cambridge.

\refs Hogerheijde M. R., van Dishoeck E. F., Blake G. A., and
van Langevelde H. J. (1998) \apj, {\em 502}, 315-336.

\refs Hogerheijde M. R., van Dishoeck E. F., Salverda J. M., and
Blake G. A (1999) \apj, {\em 513}, 350-369.

%\refs Honda, M., Kataza, H., Okamoto, Y.K., Miyata, T., Yamashita, T.,
%Sako, S., Takubo, S., and Onaka, T. (2003) \apj, {\em 585}, L59-L63.

\refs Jayawardhana R., Hartmann L.,
         Fazio G., Fisher R. S., Telesco C. M., and Pi{\~n}a R. K.
         (1999) \apj, {\em 521}, L129-L132.

\refs Jensen E. L. N. and Mathieu R. D. (1997) \aj, {\em 114}, 301-316.

\refs Keller L. P., Hony S., Bradley J. P., Molster F. J., Waters L. B. F. M., 
et al.
%Bouwman J., de Koter A., Brownlee D.E., Flynn G.J., Henning T., and  Mutschke H. 
(2002) {\em Nature}, {\em 417}, 148-150.

\refs Kemper F., Vriend W. J., and Tielens A. G. G. M. (2004) \apj, 
{\em 609}, 826-837.

\refs Kessler-Silacci J. E., Hillenbrand L. A., Blake Geoffrey A., 
and Meyer M. R. (2005) \apj, {\em 622}, 404-429.

\refs Kessler-Silacci J. E., Augereau J.-C., Dullemond C. P.,
Geers V., Lauhis F.,  et al.
%Evans N.J.II, van Dishoek E., Blake G.A., 
%Boogert A.C.A., and 4 coauthors 
(2006) \apj, in press
(astro-ph/0511092).

%\refs Krivova, N., Krivov, A.V., and Mann, I. (2000). The disk
% of $\beta$ Pictoris in the light of polarimetric data. {\em
% Astrophys. J., 539}, 424-434.

\refs Kr\"ugel E. and  Siebenmorgen R. (1994)  \aap, {\em 288}, 929-941.

\refs Leinert Ch., et al. (2006), {\em in preparation}.

\refs 
Lucas P. W., Fukagawa M., Tamura M., Beckford A. F., Itoh Y., 
et al.
%Murakawa K., Suto H., Hayashi S.S., Oasa Y., Naoi T., and 3 coauthors
(2004) 
%High-resolution imaging
%polarimetry of HL Tau and magnetic field structure. 
{\em Mon.\ Not.\ Roy.\ Astr.\ Soc., 352}, 1347-1364.

\refs Marsh K. A. and Mahoney M. J. (1992) \apj, {\em 395}, L115-L118.

\refs McCabe C., Duch\^{e}ne G., and Ghez A. M. (2003). {\em
Astrophys. J., 588}, L113-L116.

\refs McCaughrean M. J., Stapelfeldt K. R., and Close L. M. (2000)
In {\em Protostars and Planets IV} (V. Mannings, A. P.
Boss, and S. S. Russell, eds.), pp. 485-507. Univ. of Arizona, Tucson,

\refs Meeus G., Waters L. B. F. M., Bouwman J., van den Ancker M. E., 
Waelkens C., and Malfait K. (2001) \aap, {\em 365}, 476-490.

\refs Meeus G., Sterzik M., Bouwman J., and Natta A. (2003) \aap,
{\em 409}, L25-L29.

\refs Menshchikov A. B. and Henning Th. (1997) \aap, {\em 318}, 879-907.

\refs Min M., Hovenier J. W., de Koter A., Waters L. B. F. M., and Dominik C.
(2005) {\em Icarus}, {\em 179}, 158-173.

\refs Min M., Dominik C., Hovenier J. W., de Koter A., and Waters L. B. F. M.
(2006) \aap, {\em 445}, 1005-1014.

\refs Miyake K. and Nakagawa Y (1995) \apj, {\em 441}, 361-384.

\refs Muzerolle J., Calvet N., Brice{\~n}o C., Hartmann L., and  Hillenbrand L. (2000) \apj, {\em 535}, L47-L50.
%Accretion in TWHya

\refs Muzerolle J., Hillenbrand L., Calvet N., Brice{\~n}o C., and
Hartmann L. (2003) \apj, {\em 592}, 266-281.
%accretion in BDs

\refs Muzerolle J., Megeath S. T., Gutermuth R. A., Allen L. E., Pipher J. L., 
et al.
%Hartmann, L., Gordon, K. D., Padgett, D. L., Noriega-Crespo, A., Myers, P. C., and 8 coauthors 
(2004) \apjs, 154, 379-384.
% Three objects with inner gaps in NGC7129

\refs Muzerolle, J., Adame, L., D'Alessio, P., Calvet, N.,
Luhman, K. L., et al.
%Muench, A.A., Lada, C.J., Rieke, G.H., Siegler, N., 
%Trilling, D.E., Young E.T., Allen, L.,
%Hartman, L., \& Megeath, S.T., 
(2006) \apj, submitted.

% BD disks with gaps?

%\refs Natta A., Meyer M. R., and Beckwith S. V. W. (2000) \apj, {\em 534},
%838-845.

\refs Natta A. and Testi L. (2004) In
{\em Star Formation in the Interstellar Medium: In Honor of 
David Hollenbach, Chris McKee and Frank Shu}
(D. Johnstone, F.C. Adams, D.N.C. Lin, D.A. Neufeld, and E.C. Ostriker eds.),
 pp. 279-285. ASP, San Francisco.

\refs Natta A., Testi L., Neri R., Shepherd D. S., and Wilner D. J.
(2004) \aap, {\em 416}, 179-186.

\refs Natta A., Testi L., and Randich S. (2006) \aap, {\em submitted}

\refs Ossenkopf V. and  Henning Th. (1994) \aap, {\em 291}, 943-959.

\refs Peeters E., Hony S., Van Kerckhoven C., Tielens A. G. G. M., 
Allamandola L. J., Hudgins D. M., and Bauschlicher C. W.
(2002) \aap, {\em 390}, 1089-1113.

\refs  Pollack J. B., Hollenbach D., Beckwith S., Simonelli D. P., Roush T., and Fong W.  (1994) \apj, {\em 421}, 615-639.

\refs Przygodda F., van Boekel R., \'Abrah\'am P., Melnikov S. Y., 
Waters L. B. F. M., and Leinert Ch. (2003) \aap, {\em 412}, L43-L46.

\refs Qi C.,  Ho P. T. P., Wilner D. J., Takakuwa S., Hirano N.,
et al.
%Ohashi N., Bourke T.L., Zhang Q., Blake G.A., Hogerheijde M., and 3 coauthors 
(2004) \apj, {\em 616}, L11-L14. 

\refs Quillen A. C., Blackman E. G., Frank A., and Varni{\` e}re P. 
(2004) \apj, {\em 612}, L137-L140.

\refs Ressler M. E.  and Barsony M. (2003) \apj, {\em 584}, 832-842.

\refs Rice W. K. M., Wood K.,
Armitage P. J., Whitney B. A., and Bjorkman J. E. (2003) \mnras, {\em 342}, 
79-85.

\refs Rodmann J., Henning Th., Chandler C. J., Mundy L. G., and Wilner D. J.
(2006) \aap, {\em 446}, 211-223.

\refs Sargent B., Forrest W. J., D'Alessio P., Najita J., Li A.,
et al.
%Calvet, N., Furlan, E., Green, J.D., Kim, K.H., Watson, D.M.,
%Sloan, G.C., Markwick-Kemper, F., Chen, C.H., Hartmann, L., Keller, L.D.,
%Herter, T.L., Brandl, B.R., and Houck, J.R.
(2005) poster  at the IAU Symp. 231 on ``Astrochemistry Throughout the Universe: Recent Successes and Current Challenges'', Asilomar, California.

\refs Sargent, B., Forrest, W. J., D'Alessio, P., Najita, J., Li, A., 
et al.
% Calvet, N., Furlan, EL, Green, J.D., Kim, K.H., Watson, D.M., Sloan, G.C.,
%Chen, C.H., Hartmann, L., Keller, L.D., \&
%Houck, J.R., 
(2006) \apj, submitted.

%\refs 
%Sicilia-Aguilar A., Hartmann L. W., Hern\'andez J., Brice\~no C.,
%and  Calvet N.
%(2005) \aj, {\em 130}, 188-209.        
%Cepheus OB2: Disk Evolution and Accretion at 3-10 Myr ??

\refs Schr\"apler R., and Henning Th. (2004). {\em Astrophys. J., 614},
   960-978.
  
\refs Shuping R. Y., Bally J.,  Morris M., and  Throop H. (2003)
\apj, {\em 587}, L109-112.

%\refs Scott, E. R. D., and Krot, A. N. (2005). 
%%Thermal Processing of silicate dust in the solar nebula: clues from primitive chondritic matrices. 
%{\em Astrophys. J. 623}, 571-578.

\refs Sicilia-Aguilar, A., Hartmann, L., Calvet, N.,
Megeath, S. T., Muzerolle, J., et al.
%Allen, L., D'Alessio, P., Merin, B., Stauffer, J., Young. E., \& Lada, C. 
(2006) \apj, in press.

\refs  Sloan G. C.,  Keller L. D.,  Forrest W. J.,  Leibensperger E.,
Sargent B., et al.
%  Li A., ,  Najita, J.,   Watson, D.M.,  Brandl, B.R., and 10
%coauthors 
(2005) \apj, {\em 632}, 956-963.

\refs Suttner G. and Yorke H. W. (2001) \apj, {\em 551}, 461-447.

\refs Testi L., Natta A., Shepherd D. S., and Wilner D. J. (2001)
\apj, {\em 554},  1087-1094.

\refs Testi L., Natta A., Shepherd D. S., and Wilner D. J. (2003)
\aap, {\em 403}, 323-328.

\refs Throop H. B., Bally J., Esposito L. W., and McCaughrean M. J. (2001) {\em Science, 292}, 1686-1689.

\refs Uchida K. I., Calvet N., Hartmann L., Kemper F., Forrest W. J., et al.
%Watson, D. M., D'Alessio, P., Chen, C. H., Furlan, E., Sargent, B., and 10 coauthors
(2004) \apjs, {\em 154}, 439-442.

\refs van Boekel R., Min M., Leinert Ch., Waters L. B. F. M., 
Richichi A., et al.
%Chesneau, O., Dominik, C., Jaffe, W., Dutrey, A., Graser, U., and 13 coauthors 
(2004a) {\em Nature, 432}, 479-482.

\refs van Boekel R., Waters L. B. F. M., Dominik C., Dullemond C. P., 
Tielens A. G. G. M., and de Koter A. (2004b) \aap, {\em 418}, 177-184.
%Spatially and spectrally resolved 10 ?m emission in Herbig Ae/Be stars

\refs van Boekel R., Min M., Waters L. B. F. M., de Koter A., Dominik C., 
van den Ancker M. E., and  Bouwman J. (2005) \aap, {\em 437}, 189-208.
%A 10 ?m spectroscopic survey of Herbig Ae star disks: Grain growth and crystallization

\refs van Diedenhoven B., Peeters E., Van Kerckhoven C., Hony S., 
Hudgins D. M., Allamandola L. J. and Tielens A. G. G. M.
(2004) \apj, {\em 611}, 928-939.

\refs Van Kerckhoven C., Tielens A. G. G. M., and Waelkens C. (2002)
\aap, {\em  384}, 568-584.

\refs Voshchinnikov N. V. and Kr\"ugel E. (1999). 
%Circumstellar discs of $\beta$ Pictoris: constraints on grain properties from polarization. 
{\em Astron. Astrophys., 352}, 508-516.

\refs Voshchinnikov N. V., Ilin V. B., Henning Th., and Dubkova D. N. 
  (2006) \aap,  {\em 445}, 167-177.

\refs Weidenschilling S. J.  (1997)
       {\em Icarus, 127}, 290-306.

\refs Weidenschilling S. J. and Cuzzi J. N. (1993) 
%Formation of planetesimals in the solar system. 
In {\em Protostars and Planets III} (E.H. Levy and J.N. Lunine, eds.), Univ. of Arizona Press,
Tucson, 1031-1060.

\refs Weingartner J. C. and Draine B. T. (2001) \apj, {\em 548}, 296-309.

%\refs Whitney, B., Kenyon, S. J., and Gomez, M. (1997).
%%Near-infrared imaging polarimetry of embedded young stars in the
%%Taurus-Auriga molecular cloud. 
%{\em Astrophys. J.} 485, 703-xxx.

\refs Wilner D. J., Ho P. T. P., Kastner J. H. and Rodr\'\i guez L. F. (2000)
\apj, {\em 534}, L101-104.

\refs Wilner D. J., D'Alessio P., Calvet N., Claussen M. J.,
and  Hartmann L. (2005), \apj,  {\em 626},  L109-L112.

\refs Wurm G., Paraskov G., and Krauss O. (2005) {\em Icarus},
{\em 178},  253-263.
\
\end{document}